\documentclass[sigchi, nonacm=true]{acmart}

\usepackage[utf8]{inputenc}
\usepackage[ngerman]{babel}
\usepackage{graphics}
\usepackage{rotating}
\usepackage{diagbox}
\AtBeginDocument{%
  \providecommand\BibTeX{{%
    \normalfont B\kern-0.5em{\scshape i\kern-0.25em b}\kern-0.8em\TeX}}}

\setcopyright{none}
\copyrightyear{2020}
\acmYear{2020}
\acmDOI{x}

\acmConference[MuC'20 Workshops]{Proceedings of the Mensch und Computer 2020 Workshop on Usable
Security und Privacy.}{6. bis 9. Sept. 2020}{Magdeburg, Deutschland}
\acmBooktitle{MuC'20 Workshop Proceedings}
\acmPrice{0}
\acmISBN{xxx-x-xxxx-xxxx-x/xx/xx}

\usepackage{wrapfig}

\usepackage{enumitem}
\newlist{todolist}{itemize}{2}
\setlist[todolist]{label=$\square$}

\usepackage{pgfplots}\pgfplotsset{compat=1.10}
\usepackage{pgfplotstable}
\usepackage{lmodern}
\usepackage{wasysym}

\usepackage{ifthen}
\newboolean{long} 
\newboolean{blindForReview} 

\setboolean{long}{true}  
\setboolean{blindForReview}{false}
\ifthenelse{\boolean{blindForReview}}{
\newcommand{\longVersionURL}{\url{https://arxiv.org/exampleURL}}
}{
\newcommand{\longVersionURL}{\url{https://arxiv.org/submit/3257727/view}}
} 
\newcommand{\zeierAPI}{eUCRITE}
\newcommand{\anEins}{Viele Bei\-spie\-le}
\newcommand{\anNrEins}{9}%1.2.1
\newcommand{\anZwei}{Vermittelt (technisches) Hintergrundwissen}
\newcommand{\anNrZwei}{12}%2
\newcommand{\anDrei}{Gute Struktur}
\newcommand{\anNrDrei}{2}%1.1
\newcommand{\anVier}{Klassische Referenz}
\newcommand{\anNrVier}{14}%3
\newcommand{\anFuenf}{IDE Integration}
\newcommand{\anNrFuenf}{19}%5
\newcommand{\anSechs}{R{\"u}ckfragen und Kommentarfunktion}
\newcommand{\anNrSechs}{21}%6
\newcommand{\anSieben}{Verlinkungen zur Referenz / Signalisierte Text-zu-Code-Verbindungen}
\newcommand{\anNrSieben}{3}%1.1.1
\newcommand{\anAcht}{Beinhaltet die Fehlerbehandlung}
\newcommand{\anNrAcht}{11}%1.2.2
\newcommand{\anNeun}{Erweiterte Suche}
\newcommand{\anNrNeun}{4}%1.1.2
\newcommand{\anZehn}{Konzeptionelle Informationen mit verwandten Aufgaben integrieren}
\newcommand{\anNrZehn}{8}%1.2
\newcommand{\anElf}{Verschiedene Niveaus}
\newcommand{\anNrElf}{5}%1.1.3
\newcommand{\anZwoelf}{Negativbeispiele}
\newcommand{\anNrZwoelf}{10}%1.2.1.1
\newcommand{\anDreizehn}{Sourcecode}
\newcommand{\anNrDreizehn}{20}%5.1
\newcommand{\anVierzehn}{Aktualit{\"a}t}
\newcommand{\anNrVierzehn}{16}%4.1
\newcommand{\anFuenfzehn}{Vollst{\"a}ndigkeit}
\newcommand{\anNrFuenfzehn}{15}%4
\newcommand{\anSechzehn}{Suchmaschinenfreundlich}
\newcommand{\anNrSechzehn}{22}%7
\newcommand{\anSiebzehn}{Transparente Versionierung der API}
\newcommand{\anNrSiebzehn}{17}%4.2
\newcommand{\anAchtzehn}{Aufzeigen von Alternativen}
\newcommand{\anNrAchtzehn}{24}%9
\newcommand{\anNeunzehn}{{\"A}nderungshistorie der Dokumentation}
\newcommand{\anNrNeunzehn}{18}%4.3
\newcommand{\anZwanzig}{Inhalt gem{\"a}{\ss} der API-Funk\-tio\-na\-li\-t{\"a}t organisieren}%\newcommand{\anEinundzwanzig}{Konzeptionelle Informationen mit verwandten Aufgaben integrieren} Mit Anforderung 10 zusammengeführt.
\newcommand{\anNrZwanzig}{6}%1.1.4
\newcommand{\anZweiundzwanzig}{Schnelle Nutzung der API erm{\"o}glichen}
\newcommand{\anNrZweiundzwanzig}{1}%1
\newcommand{\anDreiundzwanzig}{Selektiver Zugriff auf den Code aktivieren}
\newcommand{\anNrDreiundzwanzig}{7}%1.1.5
\newcommand{\anVierundzwanzig}{Wichtige Informationen redundant bereitstellen}
\newcommand{\anNrVierundzwanzig}{23}%8
\newcommand{\anFuenfundzwanzig}{Explizite Dokumentation {\"u}ber die Performance}
\newcommand{\anNrFuenfundzwanzig}{13}%2.1

\makeatletter
\def\hlinewd#1{%
  \noalign{\ifnum0=`}\fi\hrule \@height #1 \futurelet
   \reserved@a\@xhline}
\makeatother

\newcommand{\bline}{0.8pt}

\begin{document}

%%
%% The "title" command has an optional parameter,
%% allowing the author to define a "short title" to be used in page headers.
\title{Zur Benutzbarkeit und Verwendung von API-Dokumentationen}
\ifthenelse{\boolean{blindForReview}}{
%Anonym
\author{blinded for review}
\orcid{xxx}
\affiliation{%
   \institution{}
   \streetaddress{}
   \city{}
   \state{}
   \postcode{}
}
\renewcommand{\shortauthors}{Name et al.}
}{
%Autorennamen
%%
%% The "author" command and its associated commands are used to define
%% the authors and their affiliations.
%% Of note is the shared affiliation of the first two authors, and the
%% "authornote" and "authornotemark" commands
%% used to denote shared contribution to the research.

\author{Rolf Huesmann}
\orcid{0000-0003-0798-2919}

\author{Alexander Zeier}
\orcid{0000-0003-1717-5029}

\author{Andreas Heinemann}
\orcid{0000-0003-0240-399X}

\author{Alexander Wiesmaier}
\orcid{0000-0002-1144-549X}

\email{vorname.nachname@h-da.de}
\affiliation{%
   \institution{Hochschule Darmstadt}
   \streetaddress{Schöfferstraße 3}
   \city{Darmstadt}
   \state{Deutschland}
   \postcode{64295}
}

% \author{Rolf Huesmann}
% \orcid{0000-0003-0798-2919}
% \affiliation{%
%   \institution{Hochschule Darmstadt}
%   \streetaddress{Schöfferstraße 3}
%   \city{Darmstadt}
%   \state{Deutschland}
%   \postcode{64295}
% }
% \email{rolf.huesmann@h-da.de}
% \author{Alexander Zeier}
% \orcid{0000-0003-1717-5029}
% \affiliation{%
%   \institution{Hochschule Darmstadt}
%   \streetaddress{Schöfferstraße 3}
%   \city{Darmstadt}
%   \state{Deutschland}
%   \postcode{64295}
% }
% \email{alexander.zeier@h-da.de}
% \author{Andreas Heinemann}
% \orcid{0000-0003-0240-399X}
% \affiliation{%
%   \institution{Hochschule Darmstadt}
%   \streetaddress{Schöfferstraße 3}
%   \city{Darmstadt}
%   \state{Deutschland}
%   \postcode{64295}
% }
% \email{andreas.heinemann@h-da.de}
% \author{Alexander Wiesmaier}
% \orcid{0000-0002-1144-549X}
% \affiliation{%
%   \institution{Hochschule Darmstadt}
%   \streetaddress{Schöfferstraße 3}
%   \city{Darmstadt}
%   \state{Deutschland}
%   \postcode{64295}
% }
% \email{alexander.wiesmaier@h-da.de}

%%
%% By default, the full list of authors will be used in the page
%% headers. Often, this list is too long, and will overlap
%% other information printed in the page headers. This command allows
%% the author to define a more concise list
%% of authors' names for this purpose.
\renewcommand{\shortauthors}{Huesmann et al.}
}
%%
%% The abstract is a short summary of the work to be presented in the
%% article.
\begin{abstract}
Eine gute Dokumentation ist essenziell für eine gute Benutzbarkeit von (Sicherheits-)APIs, d.h. insbesondere für die korrekte Verwendung der APIs. Anforderungen an eine gute Dokumentation von APIs wurden in mehreren Arbeiten beschrieben, jedoch gibt es bislang keine technische Umsetzung (im folgenden Dokumentationssystem genannt), welche diese Anforderungen umsetzt. Die Anforderungen lassen sich unterteilen in Anforderungen an das Dokumentationssystem und Anforderungen an den Dokumentationsinhalt. Aus $13$ identifizierten Anforderungen an ein Dokumentationssystem selbst wurden im Rahmen dieser Arbeit $9$ in einen Prototypen umgesetzt und in einer Nutzerstudie mit 22 Probanden unter Verwendung einer kryptografischen API evaluiert. Es hat sich gezeigt, dass  die Umsetzung der Anforderung \emph{\anZweiundzwanzig} zum einen wesentlich von der Qualität der eingepflegten Inhalte abhängt, zum anderen aber auch 5 weitere der betrachteten Anforderungen bzw. deren Umsetzungen subsumiert. Die zwei weiteren umgesetzten Anforderungen (\emph{\anVier} und \emph{\anSechs}) wurden von den Probanden kaum oder nicht eingesetzt. Deren Nützlichkeit und Relevanz sollte in einer Langzeitstudie untersucht werden.
\end{abstract}

%%
%% The code below is generated by the tool at http://dl.acm.org/ccs.cfm.
%% Please copy and paste the code instead of the example below.
%%
\begin{CCSXML}
<ccs2012>
   <concept>
       <concept_id>10003120.10003121.10003122.10010854</concept_id>
       <concept_desc>Human-centered computing~Usability testing</concept_desc>
       <concept_significance>500</concept_significance>
       </concept>
   <concept>
       <concept_id>10003120.10003121.10003122.10003334</concept_id>
       <concept_desc>Human-centered computing~User studies</concept_desc>
       <concept_significance>500</concept_significance>
       </concept>
   <concept>
       <concept_id>10002978.10003029.10011703</concept_id>
       <concept_desc>Security and privacy~Usability in security and privacy</concept_desc>
       <concept_significance>500</concept_significance>
       </concept>
 </ccs2012>
\end{CCSXML}

\ccsdesc[500]{Human-centered computing~Usability testing}
\ccsdesc[500]{Human-centered computing~User studies}
\ccsdesc[500]{Security and privacy~Usability in security and privacy}

%%
%% Keywords. The author(s) should pick words that accurately describe
%% the work being presented. Separate the keywords with commas.
\keywords{API Dokumentation, Dokumentationssystem, Benutzbarkeit, Gebrauchstauglichkeit, Nutzerstudie, \zeierAPI, Tink}

%%
%% This command processes the author and affiliation and title
%% information and builds the first part of the formatted document.
\maketitle

%-----------------------------------------------------------------------------------------------------------------
\section{Einleitung und Motivation}
\label{sec:Einleitung}

 Mehrere Studien \cite{zeier_api_2019,nadi_jumping_2016,zibran_useful_2011,stylos_implications_2008,stylos_case_2008,heydarnoori_two_2012,montandon_documenting_2013,subramanian_live_2014} haben die Bedeutung von Dokumentationen für die Benutzbarkeit einer API herausgestellt. 
Weitere Arbeiten \cite{huesmann_eigenschaften_2019,meng_how_2019,meng_application_2018,robillard_field_2011} haben gezeigt, was eine gute bzw. gut benutzbare Dokumentation ausmacht. 
Viele API-Dokumentationen erfüllen diese Anforderungen jedoch nicht oder nicht ausreichend \cite{acar_developers_2017}, was zu einer fehlerhaften Verwendung der API führt und insbesondere bei Krypto-APIs die intendierten Sicherheitsziele einer Anwendung gefährdet. Bei einer unzureichenden Benutzbarkeit von API-Dokumentationen wird durch den Entwickler häufig auf andere Quellen wie beispielsweise die \emph{Question and Answer} (Q\&A) Plattform StackOverflow\footnote{\url{https://stackoverflow.com}, Abgerufen am 7.10.2019} zurückgegriffen. Jedoch sind hier die Codebeispiele oft nicht fehlerfrei \cite{fischer_stack_2017}. 

Softwareentwickler sind Schlüsselpersonen, da sie Code produzieren, welcher mög\-lich\-er\-weise auf Millionen von Geräten ausgeführt wird. Dadurch werden sie zu Multiplikatoren von Sicherheitsrisiken. Gerade im Bereich der kryptografischen APIs kann ein Programmierfehler eine große Tragweite haben \cite{wijayarathna_generic_2017}. Deswegen sollte besonders auf die Verständlichkeit und Benutzbarkeit von kryptografischen API-Dokumentationen für Softwareentwickler geachtet werden. 

In dieser Arbeit wird ein Dokumentationssystem für kryptografische APIs anhand existierender Anforderungen entwickelt, prototypisch implementiert und in einer Nutzerstudie hinsichtlich der Benutzbarkeit eva\-luiert.
Die Anforderungen wurden durch eine Literaturrecherche 
\ifthenelse{\boolean{blindForReview}}{
%Anonym
von uns
}{
%Autorennamen
\cite{huesmann_konzeption_2020}
} 
ermittelt und anhand der Ab\-hängig\-keiten zueinander geordnet (siehe Abschnitt~\ref{sec:anforderungen}). Anschließend wird der entwickelte Prototyp (Abschnitt~\ref{sec:Prototyp}) und die durchgeführte Nutzerstudie (Abschnitt~\ref{sec:Studie}) sowie deren Ergebnisse (Abschnitt~\ref{sec:Ergebnisse}) beschrieben. 
In Abschnitt~\ref{sec:Limitierung} werden Limitierungen dieser Arbeit aufgezeigt.
Abschließend gibt es eine Zusammenfassung und einen Ausblick (Abschnitt~\ref{sec:Zusammenfassung}). Folgender Abschnitt~\ref{sec:anforderungen} geht auf die Anforderungsanalyse ein.

\section{Anforderungsanalyse anhand verwandter Arbeiten} \label{sec:anforderungen}

In einer explorativen Studie haben 
\ifthenelse{\boolean{blindForReview}}{
%Anonym
Huesmann et al.
}{
%Autorennamen
die Autoren dieses Beitrags
} 
 \cite{huesmann_eigenschaften_2019} $10$ Kriterien identifiziert, die eine gut benutzbare API-Dokumentation erfüllen sollte. Dazu wurden in mehreren Fokusgruppen mit insgesamt 26 Teilnehmern die Vor- und Nachteile bestehender Dokumentationsarten und -systeme von APIs erarbeitet. 
Die Autoren Meng et al. \cite{meng_application_2018} haben mit $17$ Entwicklern jeweils ein Interview mit einem Fragenkatalog bestehend aus $45$ Fragen durchgeführt. Aus den Antworten haben sie $13$ Eigenschaften, welche im Bezug zu einer guten API-Dokumentation erwähnt wurden, extrahieren können.
Ein Jahr später haben Meng et al. \cite{meng_how_2019} die API der shipcloud GmbH\footnote{\url{https://www.shipcloud.io/ } Abgerufen am: 07.10.2019} auf ihre Benutzbarkeit untersucht. $11$ Teilnehmer haben dazu eine Programmieraufgabe mit der API gelöst und wurden anschließend mit einem Fragebogen über die Qualität der API-Dokumentation befragt. Dabei wurden $11$ Anforderungen diskutiert. 
Robillard \cite{robillard_field_2011} haben $2011$ in ihrer Arbeit $6$ Anforderungen an API-Dokumentationen beschrieben. Dazu haben sie mit $440$ Microsoft-Ent\-wick\-lern über eine explorative qualitative Studie und eine Umfrage die Benutzbarkeit einer API-Dokumentation untersucht.

Die in diesen Arbeiten genannten Anforderungen bilden die Ausgangslange für diese Arbeit. Sie sind in Tabelle \ref{tab:UebersichtDerKategorien}
zusammengefasst und nach Abhängigkeiten zueinander gruppiert. Beispielsweise trägt die Umsetzung der An\-for\-der\-ung \anNrDrei\ \emph{\anDrei}  zur Umsetzung der Anforderung \anNrZweiundzwanzig\ \emph{\anZweiundzwanzig}  bei. Diese Abhängigkeiten sind als gerichteter Graph zu verstehen.
Anforderungen, die nicht selbsterklärend sind, werden im Folgenden kurz erläutert:
\newcolumntype{L}[1]{>{\raggedright\arraybackslash}p{#1}} % linksbündig mit Breitenangabe
\newcolumntype{C}[1]{>{\centering\arraybackslash}p{#1}} % zentriert mit Breitenangabe
\newcolumntype{R}[1]{>{\raggedleft\arraybackslash}p{#1}} % rechtsbündig mit Breitenangabe
\ifthenelse{\boolean{long}}{
%long version
\begin{table*}
	\centering
	   % \begin{tabular}{|l|L{8cm}|l|c|c|c|c|}
	    
		\begin{tabular}{|l|L{10,8cm}|L{1,8cm}|c|c|c|c|}
			\hline
			$\#$ & \textbf{Anforderung} & Abhängig von &
			{
			\begin{sideways} 
			    \cite{huesmann_eigenschaften_2019} 
			\end{sideways}
			}
			&
			{
			\begin{sideways} 
			    \cite{meng_application_2018} 
			\end{sideways}
			}
			 &
			{
			\begin{sideways} 
 		    \cite{meng_how_2019} 
			\end{sideways}
			}
			 &
			{
			\begin{sideways} 
			    \cite{robillard_field_2011}
			\end{sideways}
			}
			 \\
			\hline
			\anNrZweiundzwanzig & \anZweiundzwanzig & \anNrDrei, \anNrZehn & & & $\circ \bullet$ & \\ %& SChnelle Nutzung
			\hline
			\anNrDrei & \anDrei & \anNrSieben, \anNrNeun, \anNrElf, \anNrZwanzig, \anNrDreiundzwanzig  & $\circ \bullet$ & & $\circ \bullet$ & $\circ \bullet$ \\ %Gute Strucktur
			\hline
			\anNrSieben & \anSieben & & $\bullet$ & & $\bullet$ & \\ %Verlinkung zur refernez
			\hline
			\anNrNeun & \anNeun & & & $\bullet$ & $\bullet$ & \\ %& Erweiterte Such
			\hline
			\anNrElf & \anElf & & $\circ$ & & & \\ %& Verschiedene Niveaus
			\hline
			\anNrZwanzig & \anZwanzig & & & & $\circ$ & \\%& Inhalt API-Funktionalität
			\hline
			\anNrDreiundzwanzig & \anDreiundzwanzig & & & & $\bullet$ & \\%& Selektiver Code zugriff
			\hline
			\anNrZehn & \anZehn & \anNrEins, \anNrAcht & & & $\circ$ & $\circ$ \\%& Konzeptionellel Informationen
			\hline
			\anNrEins & \anEins & \anNrZwoelf & $\circ$ & $\circ$ & $\circ$ & $\circ$ \\%& Viele Beispiele
			\hline
			\anNrZwoelf & \anZwoelf & & $\circ$ & & & \\%& Negativbeispiele
			\hline
			\anNrAcht & \anAcht & & & $\circ$ & & $\circ$ \\%& Fehlerbehandlung
			\hlinewd{\bline}
			\anNrZwei & \anZwei & \anNrFuenfundzwanzig & $\circ$ & $\circ$ & $\circ$ & $\circ$ \\%& Hintergundwissen
			\hline
			\anNrFuenfundzwanzig & \anFuenfundzwanzig & & & & & $\circ$ \\%& Perfomace
			\hlinewd{\bline}
			\anNrVier & \anVier & & $\circ \bullet$ & $\circ \bullet$ & & \\%& Klassiche Referenz
			\hlinewd{\bline}
			\anNrFuenfzehn  & \anFuenfzehn & \anNrVierzehn, \anNrSiebzehn, \anNrNeunzehn  & & $\circ \bullet$ & & \\%& Vollständigkeit
			\hline
			\anNrVierzehn & \anVierzehn & & & $\circ$ & & \\%& Aktualität
			\hline
			\anNrSiebzehn & \anSiebzehn & & & $\circ \bullet$ & & \\%& Transparente Versionierung
			\hline
			\anNrNeunzehn & \anNeunzehn & & & $\bullet$ & & \\%& Änderungshistory Doku
			\hlinewd{\bline}
			\anNrFuenf & \anFuenf & \anNrDreizehn & $\circ \bullet$ & $\circ \bullet$ & & \\%& IDE Integrtion
			\hline
			\anNrDreizehn & \anDreizehn & & $\circ \bullet$ & & & \\%& SourceCode
			\hlinewd{\bline}
			\anNrSechs & \anSechs & & $\bullet$ & $\bullet$ & & \\%& Rückfragen und kOmmentare
			\hlinewd{\bline}
			\anNrSechzehn & \anSechzehn & & & $\bullet$ & & \\%& Suchmaschienen freundlich
			\hlinewd{\bline}
			\anNrVierundzwanzig & \anVierundzwanzig & & & & $\circ$ & \\%& Redudante informationen
			\hlinewd{\bline}
			\anNrAchtzehn & \anAchtzehn & & & $\circ$ & & \\%& Alternative aufzeigen
			\hline
%			21 & \anEinundzwanzig & & & $\bullet$ & \\
%			\hline
		\end{tabular}
		
	\vspace*{1ex}
	\caption{Übersicht der Anforderungen an den Dokumentationsinhalt ($\circ$) und das Dokumentationssystem ($\bullet$) einer API. Gruppiert nach Abhängigkeiten zueinander.
	}
	\label{tab:UebersichtDerKategorien}
\end{table*}
}{
%short version \longVersionURL 
\begin{table*}
	\centering
	   % \begin{tabular}{|l|L{8cm}|l|c|c|c|c|}
	    \resizebox{1.8\columnwidth}{!}{%
		\begin{tabular}{|l|L{10,8cm}|l|c|c|c|c|}
			\hline
			$\#$ & \textbf{Anforderung} & Abhängig von &
%			{
%			\begin{sideways} 
			    \cite{huesmann_eigenschaften_2019} 
%			\end{sideways}
%			}
			&
%			{
%			\begin{sideways} 
			    \cite{meng_application_2018} 
%			\end{sideways}
%			}
			 &
%			{
%			\begin{sideways} 
			    \cite{meng_how_2019} 
%			\end{sideways}
%			}
			 &
%			{
%			\begin{sideways} 
			    \cite{robillard_field_2011}
%			\end{sideways}
%			}
			 \\
			\hline
			\anNrZweiundzwanzig & \anZweiundzwanzig & \anNrDrei, \anNrZehn & & & $\circ \bullet$ & \\ %& SChnelle Nutzung
			\hline
			\anNrDrei & \anDrei & \anNrSieben, \anNrNeun, \anNrElf, \anNrZwanzig, \anNrDreiundzwanzig  & $\circ \bullet$ & & $\circ \bullet$ & $\circ \bullet$ \\ %Gute Strucktur
			\hline
			\anNrSieben & \anSieben & & $\bullet$ & & $\bullet$ & \\ %Verlinkung zur refernez
			\hline
			\anNrNeun & \anNeun & & & $\bullet$ & $\bullet$ & \\ %& Erweiterte Such
			\hline
			\anNrElf & \anElf & & $\circ$ & & & \\ %& Verschiedene Niveaus
			\hline
			\anNrZwanzig & \anZwanzig & & & & $\circ$ & \\%& Inhalt API-Funktionalität
			\hline
			\anNrDreiundzwanzig & \anDreiundzwanzig & & & & $\bullet$ & \\%& Selektiver Code zugriff
			\hline
			\anNrZehn & \anZehn & \anNrEins, \anNrAcht & & & $\circ$ & $\circ$ \\%& Konzeptionellel Informationen
			\hline
			\anNrEins & \anEins & \anNrZwoelf & $\circ$ & $\circ$ & $\circ$ & $\circ$ \\%& Viele Beispiele
			\hline
			\anNrZwoelf & \anZwoelf & & $\circ$ & & & \\%& Negativbeispiele
			\hline
			\anNrAcht & \anAcht & & & $\circ$ & & $\circ$ \\%& Fehlerbehandlung
			\hlinewd{\bline}
			\anNrZwei & \anZwei & \anNrFuenfundzwanzig & $\circ$ & $\circ$ & $\circ$ & $\circ$ \\%& Hintergundwissen
			\hline
			\anNrFuenfundzwanzig & \anFuenfundzwanzig & & & & & $\circ$ \\%& Perfomace
			\hlinewd{\bline}
			\anNrVier & \anVier & & $\circ \bullet$ & $\circ \bullet$ & & \\%& Klassiche Referenz
			\hlinewd{\bline}
			\anNrFuenfzehn  & \anFuenfzehn & \anNrVierzehn, \anNrSiebzehn, \anNrNeunzehn  & & $\circ \bullet$ & & \\%& Vollständigkeit
			\hline
			\anNrVierzehn & \anVierzehn & & & $\circ$ & & \\%& Aktualität
			\hline
			\anNrSiebzehn & \anSiebzehn & & & $\circ \bullet$ & & \\%& Transparente Versionierung
			\hline
			\anNrNeunzehn & \anNeunzehn & & & $\bullet$ & & \\%& Änderungshistory Doku
			\hlinewd{\bline}
			\anNrFuenf & \anFuenf & \anNrDreizehn & $\circ \bullet$ & $\circ \bullet$ & & \\%& IDE Integrtion
			\hline
			\anNrDreizehn & \anDreizehn & & $\circ \bullet$ & & & \\%& SourceCode
			\hlinewd{\bline}
			\anNrSechs & \anSechs & & $\bullet$ & $\bullet$ & & \\%& Rückfragen und kOmmentare
			\hlinewd{\bline}
			\anNrSechzehn & \anSechzehn & & & $\bullet$ & & \\%& Suchmaschienen freundlich
			\hlinewd{\bline}
			\anNrVierundzwanzig & \anVierundzwanzig & & & & $\circ$ & \\%& Redudante informationen
			\hlinewd{\bline}
			\anNrAchtzehn & \anAchtzehn & & & $\circ$ & & \\%& Alternative aufzeigen
			\hline
%			21 & \anEinundzwanzig & & & $\bullet$ & \\
%			\hline
		\end{tabular}
		}
	\vspace*{1ex}
	\caption{Übersicht der Anforderungen an den Dokumentationsinhalt ($\circ$) und das Dokumentationssystem ($\bullet$) einer API. Gruppiert nach Abhängigkeiten zueinander.
	}
	\label{tab:UebersichtDerKategorien}
\end{table*}
}
\begin{description}%[leftmargin=0.5cm]
    \item[\anNrZweiundzwanzig\ \anZweiundzwanzig:]
	Dies sollte \hfill \\ zum Bei\-spiel durch einen Schnelleinstieg oder einer Try\-out-Funk\-tio\-na\-li\-tät\footnote{Beispiel für eine Tryout-Funk\-tio\-na\-li\-tät im Web: Der Besucher kann Code auf der Webseite \textit{"`compilieren"'} lassen und sieht sofort ob der Code fehlerfrei oder fehlerhaft ist.} möglich sein. 
	\item[\anNrElf \ \anElf:]
	Je nach Entwicklertyp  und Wissensstand ist für ein Entwickler ein anderes Informationsangebot hilfreich. Ein Entwickler, der bspw.\ nur wenig Erfahrung mit kryptografischen APIs hat, braucht eine andere Detailtiefe und Herangehensweise als ein Entwickler, der täglich mit kryptografischen APIs arbeitet.
	\item[\anNrZwanzig \ \anZwanzig:]\hfill \\
	Die API-Dokumentation sollte nach Kategorien strukturiert sein, welche die Funktionalität oder den Inhaltsbereich der API widerspiegeln. Stößt ein Entwickler auf ein Problem, sollte er ungefähr wissen, in welchem Bereich oder welcher Methode dieses Problem verortet ist. Die Dokumentation sollte so strukturiert sein, dass der Entwickler diesen Ort in der Dokumentation gut finden kann.
	\item[\anNrDreiundzwanzig \ \anDreiundzwanzig:]
	Die Bei-\\ spiele sollten über eine geeignete Designstrategie klar vom Text zu unterscheiden sein, wodurch es für Entwickler mit einer opportunistischen Arbeitsweise nach Clarke \cite{clarke_what_2007} einfacher wird, direkt zu relevanten Beispielen zu springen.
	\item[\anNrZehn]\textbf{\anZehn:}
	Konzeptionelle Informationen verdeutlichen die Funktionsweise sowie Struktur der API und sollten zusammen mit Aufgaben, geeigneten Beispielen oder Nutzungsszenarien präsentiert werden.
	\item [\anNrVier \ \anVier:]
	In der jede Klasse, Methode und Funktion mit allen Parametern %hierarchisch geordnet 
	beschrieben wird.
    \item[\anNrVierundzwanzig\ \anVierundzwanzig:]\hfill \\
	Ent\-wick\-ler mit einer op\-por\-tunistischen Arbeitsweise sind beim Erlernen einer neuen API stärker auf Beispiele fixiert. Das bringt das Risiko mit sich, dass sie Abschnitte in der API-Dokumentation überspringen, in denen kritische konzeptionelle Informationen präsentiert werden. Dies erfordert einen Ansatz, der diese Inhalte redundant darstellt, z.B. in dem beschreibenden Dokumentationstext und zusätzlich in dem Beispiel als Codekommentar.
	\item[\anNrAchtzehn\ \anAchtzehn:]
	Die API-Doku\-men\-taion sollte alternative Methoden, Funktionen, Klassen und Parameter aufzeigen, welche beispielsweise ähnliche Funktionalitäten bieten und somit ein Problem ggfs. effizienter lösen.
\end{description}
Eine Dokumentation besteht aus zwei wesentlichen Faktoren: 1.) Aus dem Dokumentationssystem, der Plattform oder dem Framework, mit dem die Dokumentation erstellt wird. 2.) Aus den Informationen, die der Dokumentationsersteller in dem System pflegt.
Alle oben genannten Anforderungen kann das Dokumentationssystem alleine nicht erfüllen. Ein Teil davon hängt von der redaktionellen Pflege durch den Dokumentationsersteller, also dem Inhalt, ab (in Tabelle \ref{tab:UebersichtDerKategorien} durch ein $\circ$ markiert).

%  Qualität der eingepflegten Inhalte
Bei genauerer Betrachtung der Abhängigkeiten von Anforderung \anNrZweiundzwanzig\ \emph{\anZweiundzwanzig} fällt auf, dass 8 der 11 Anforderungen von der inhaltlichen Qualität der eingepflegten Informationen abhängig sind. %(In Untergruppe \anNrZehn\ alle).
Je nach Anwendungsfall sind einige dieser Anforderungen ggfs. nicht relevant.
So verfügt beispielsweise nicht jede API über verschiedene Versionen oder über ähnliche, alternative Funktionsaufrufe oder Klassen.
\ifthenelse{\boolean{long}}{
%long version
Die inhaltlichen Anforderungen, an die sich ein Dokumentationsersteller halten sollte, 
um eine optimierte Dokumentation zu erstellen, 
sind in Form einer Checkliste (siehe Anhang \ref{sec:Anhang:checkliste}) aufbereitet worden. Mit Hilfe dieser Checkliste wurden die Inhalte der im Prototyp verwendeten \zeierAPI-API erstellt. 
%was ist mit dem 'Konzept' ? -> Anhang \ref{sec:Anhang:konzept}
}{
%short version \longVersionURL 

Die inhaltlichen Anforderungen, an die sich ein Dokumentationsersteller halten sollte, um eine optimierte Dokumentation zu erstellen, sind in Form einer Checkliste\footnote{Die Checkliste und das Konzept ist in einer Langversion der Arbeit zu finden: \longVersionURL} aufbereitet worden. Mit Hilfe dieser Checkliste wurden die Inhalte der im Prototyp verwendeten \zeierAPI-API-Dokumentation erstellt. 
}
	
Aus den Anforderungen an das Dokumentationssystem (in Tabelle \ref{tab:UebersichtDerKategorien} durch ein $\bullet$ markiert) wurde ein
\ifthenelse{\boolean{long}}{
%long version
Konzept für eine webbasierte Realisierung erstellt. Dieses wird im Anhang \ref{sec:Anhang:konzept} genauer beschrieben.
}{
%short version \longVersionURL
Konzept\footnotemark[4]
%\footnote{Das Konzept ist ebenfalls in einer Langversion dieser Arbeit zu finden: \longVersionURL}
für eine webbasierte Realisierung erstellt.
} 
Anhand dessen wurde der im folgenden Abschnitt beschriebene Prototyp umgesetzt. 	
	
\section{Prototyp}
\label{sec:Prototyp}

Die Anforderungen an das Dokumentationssystem aus Abschnitt \ref{sec:anforderungen} wurden weitgehend prototypisch umgesetzt. Als zu dokumentierende API wurde die von Zeier et al. \cite{zeier_api_2019} entwickelte \zeierAPI-API gewählt, da diese bereits gute Ergebnisse zur Benutzbarkeit hervorgebracht hat. Dadurch sollte weitestgehend ausgeschlossen werden, dass negative Ergebnisse dieser Studie auf die API anstatt auf das Dokumentationssystem zurückzuführen sind. 
Die \zeierAPI-API richtet sich an App-Entwickler, die keine oder nur wenige kryptografische Kenntnisse besitzen. Der aktuelle Stand der API unterstützt kryptografische Operatoren zur Verschlüsselung und Signierung von Daten. Über Templates können Entwickler die aktuellen Verschlüsselungsparameter oder Algorithmen wählen, welche in Zukunft regelmäßig durch automatische Updates aktualisiert werden sollen.
Nachfolgender Abschnitt geht auf Anforderungen ein, die nicht oder nicht vollständig umgesetzt werden konnten.

Anforderung \anNrFuenf\ der \emph{\anFuenf} wurde nicht realisiert, da die Probanden bewusst eine separate Online-Dokumentation nutzen sollten.  
Die meisten modernen IDEs bieten eine Javadoc-Integration\footnote{Javadoc: \url{http://www.oracle.com/technetwork/java/javase/documentation/javadoc-137458.html} Abgerufen am 08.09.2019}. Durch das Einfügen von Links auf unserem Dokumentationssystem in das Javadoc, kann eine erste IDE-Integration unseres Dokumentationssystems erreicht werden.
Anforderung \anNrDreizehn\ wurde nicht betrachtet, da der Source-Code der \zeierAPI-API in der IDE des Entwicklers angezeigt wurde. 
Die \emph{Suchmaschinenfreundlichkeit} (Anforderung \anNrSechzehn) wurde in dieser Arbeit außen vorgelassen, da die Erreichbarkeit der \zeierAPI-API-Dokumentation in Suchmaschinen für diese Studie nicht gegeben ist, da die API bisher nicht öffentlich verfügbar gemacht wurde. 
Da der Prototyp auf dem Wordpress\footnote{Wordpress: \url{https://de.wordpress.org} Abgerufen am 16.12.2019} Content Management System basiert, wäre ein Auffinden durch Suchmaschinen jedoch leicht realisierbar, da hier Search Engine Optimization Plugins einfach integriert werden können.
Anforderung \anNrSiebzehn\ wurde durch einen Versions-Schalter im Footer der Seite angedeutet. Dieser wurde allerdings nicht implementiert, da die \zeierAPI-API bisher nicht über unterschiedliche Versionen verfügt. Aus demselben Grund wurde die \emph{\anNeunzehn} (Anforderung \anNrNeunzehn) nicht realisiert.
Trotz diesen Restriktionen haben wir uns für die \zeierAPI-API entschieden, da zum Zeitpunkt des Studiendesings laut TIOBE Index\footnote{TIOBE Index: \url{https://www.tiobe.com/tiobe-index/} Abgerufen am: 01.08.2019} Java die meistbenutzte Programmiersprache war und uns bisher keine andere Java-Api bekannt ist, welche bereits auf Benutzbarkeit überprüft wurde. 

Wenden wir uns nun den Anforderungen zu, die durch unsere Implementierung erfüllt werden. 

 \ifthenelse{\boolean{long}}{
%long version
\begin{figure*}
 \centering
\includegraphics[width=.6\textwidth]{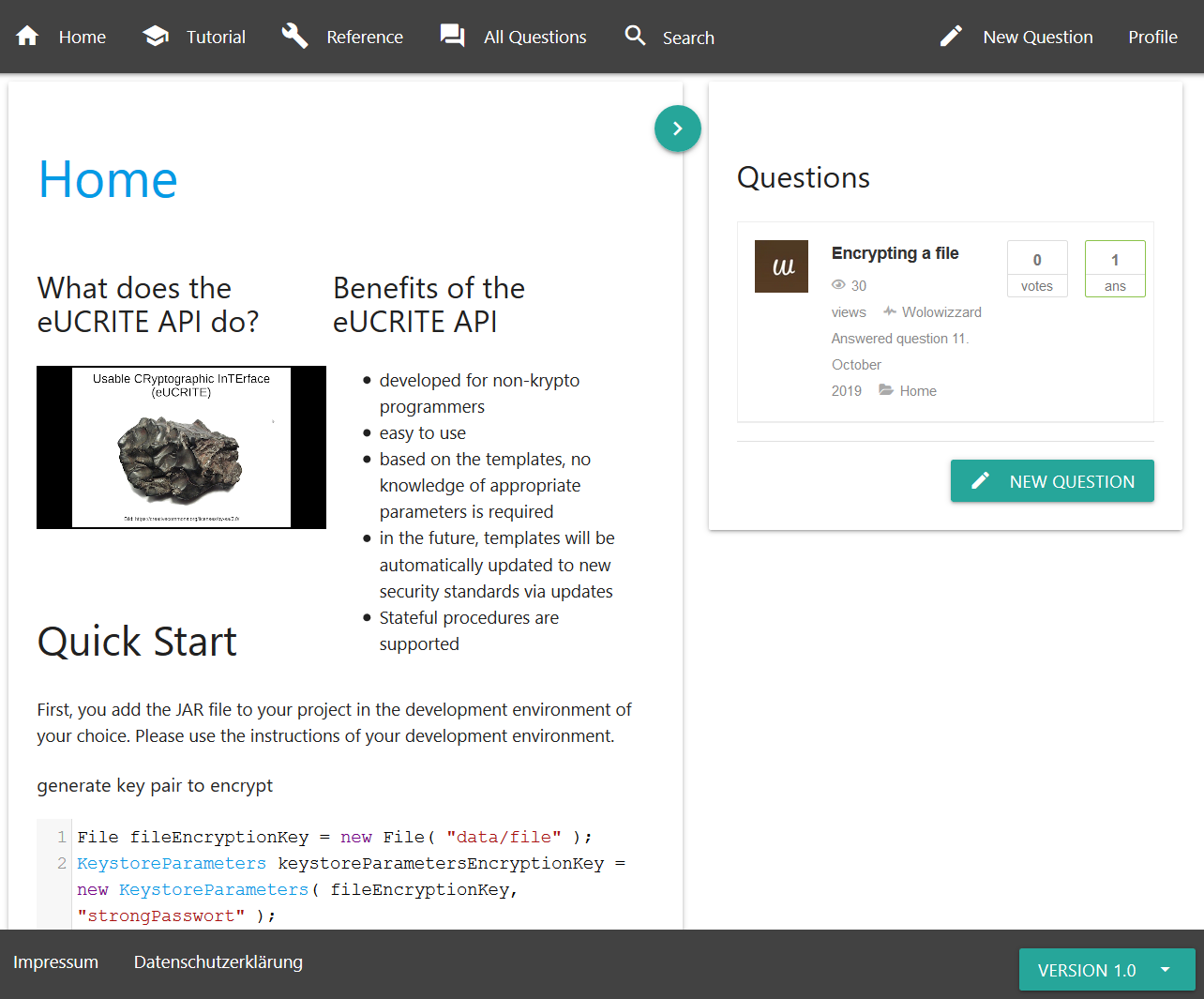}
 \caption{Startseite der \zeierAPI-API-Dokumentation}
 \label{fig:Vorgenhensweise:NextDoku-Home}
\end{figure*}
}{
%short version \longVersionURL
\begin{figure}
 \centering
\includegraphics[width=.4\textwidth]{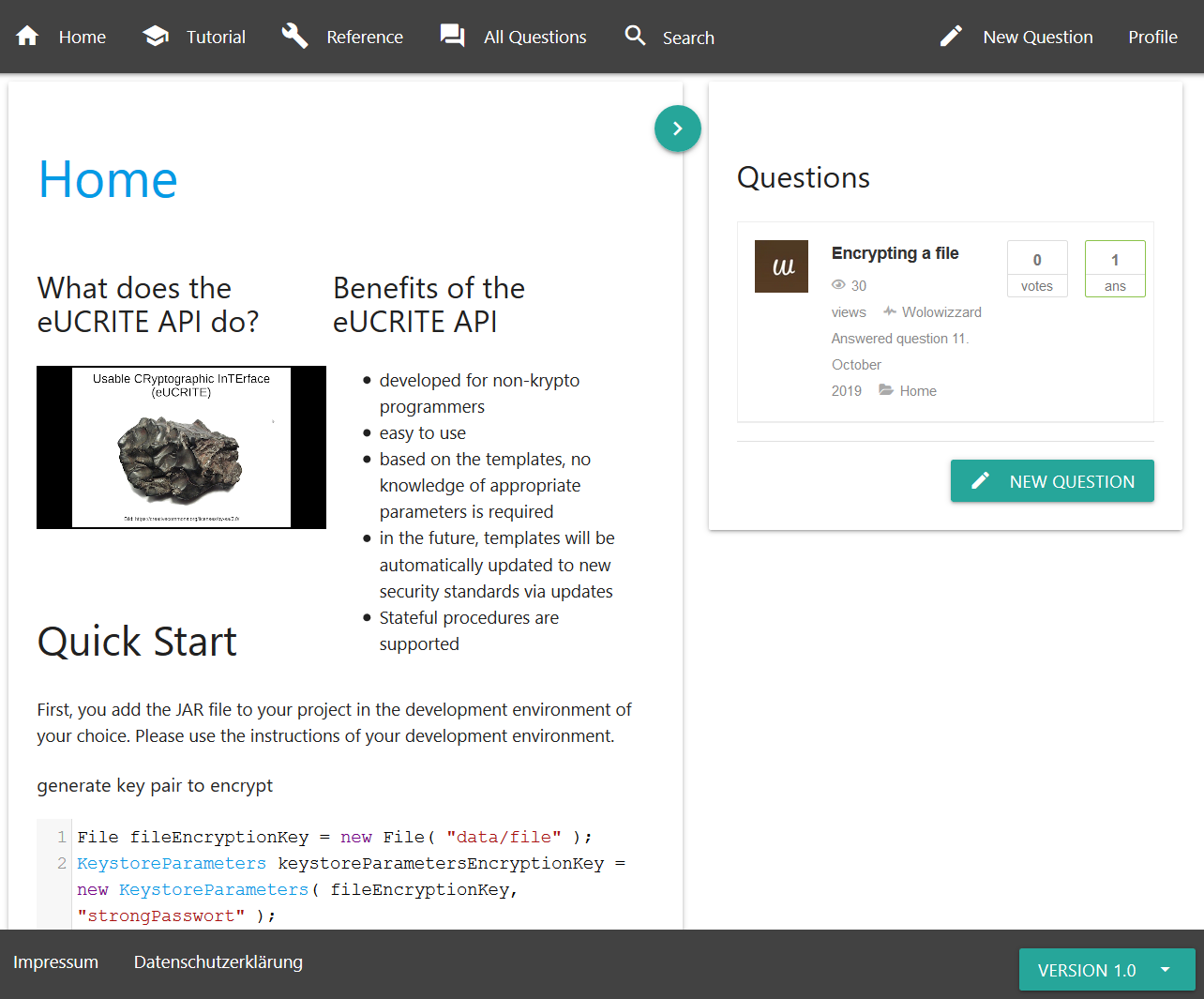}
 \setlength{\belowcaptionskip}{-14pt}
 \caption{Startseite der \zeierAPI-API-Dokumentation}
 \label{fig:Vorgenhensweise:NextDoku-Home}
\end{figure}
}
Abbildung \ref{fig:Vorgenhensweise:NextDoku-Home} zeigt die Startseite der prototypischen Dokumentation. Der Fragen-Bereich (Anforderung~\anNrSechs) ist auf der rechten Seite zu sehen. Dieser zeigt die Fragen an, die für die aktuell angezeigte Seite relevant sind. Der Besucher kann so direkt thematisch passend zum Inhalt der Dokumentation eine Frage stellen, Antworten lesen und verfassen.
Anforderung \anNrDrei\ \emph{\anDrei} wird durch ein auf das Wesentliche reduziertes und mit Icons optisch aufbereitetes Header-Menü realisiert. In ihm ist die Suche integriert, welche auf der Trefferseite verfeinert werden kann und so Anforderung \anNrNeun\ entspricht.
Unter dem Menü-Punkt "`Reference"' ist eine \emph{klassische Referenz} (Anforderung \anNrVier) angegliedert. Diese könnte zukünftig automatisiert von dem Dokumentationssystem aus dem Javadoc der API generiert und durch den Dokumentationsersteller mit wichtigen Informationen verfeinert werden.
Anforderung~\anNrSieben\ wurde durch die Verlinkung von Klassen-Namen und -Aufrufen in Texten und Code-Beispielen realisiert. Die Besucher gelangen durch das Klicken auf einen Klassen-Aufruf in Code-Beispielen in die jeweilige Klassen-Beschreibung der API-Referenz.
Code-Beispiele werden mit Syntaxhervorhebung, Zeilennummern und Einrückungen optisch von Beschreibungen hervorgehoben. Damit wurde Anforderung \anNrDreiundzwanzig\ umgesetzt.

Die inhaltlichen Anforderungen (in Tabelle \ref{tab:UebersichtDerKategorien} durch ein $\circ$ markiert) wurden vom Autor der \zeierAPI-Dokumen\-tation 
\ifthenelse{\boolean{long}}{
%long version
anhand der im Anhang \ref{sec:Anhang:checkliste} beigelegten Checkliste 
}{
%short version \longVersionURL 
anhand der Checkliste %\footnote{Die Checkliste ist in einer Langversion der Arbeit zu finden: \longVersionURL} 
}umgesetzt.

\section{Studiendesign und Durchführung}
\label{sec:Studie}
\ifthenelse{\boolean{long}}{
%long version
\begin{table*}
	\centering
	    
		\begin{tabular}{|l|L{6.5cm}|c|c|c|c|c|c|c|c|}
			\hline
			$\#$ & {
			\diagbox[width=6.5cm,innerleftsep=.5cm,innerrightsep=0pt]
			{\textbf{Anforderung}}
			{ 
			%\begin{sideways} 
			    \textbf{Methode}  
			%\end{sideways}
			}
			} & {
			\begin{sideways} \raggedleft  
			    Q\&A Fragen  
			\end{sideways}
			} & {
			\begin{sideways} 
			    Aussage 1 %supported me in solving the task 
			\end{sideways}
			} & {
			\begin{sideways} 
			    Aussage 2 %Found my way esaily 
			\end{sideways}
			} & {
			\begin{sideways} 
			    Aussage 3 %by accessing, i found useful help  
			\end{sideways}
			} & {
			\begin{sideways} 
			    Aussage 4 %helpful expations
			\end{sideways}
			} & {
			\begin{sideways} 
			    Aussage 5 %helpful code  
			\end{sideways}
			} & {
			\begin{sideways} 
			    Eye-Tracking  
			\end{sideways}
			} & {
			\begin{sideways}
			    Benötigte Zeit
			\end{sideways}
			} \\
			\hline
			\anNrZweiundzwanzig & \anZweiundzwanzig & & & & $\star$ & & & & $\star\star$ \\ %& 1
			\hline
			\anNrDrei & \anDrei & & $\star$ & $\star\star$ & $\star$ & & & & \\ %& 2
			\hline
			\anNrSieben & \anSieben & & & & & & & $\star\star$ & \\ %& 3 
			\hline
			\anNrNeun & \anNeun & & & & & & & $\star\star$ & \\ %& 4 
			\hline
			\anNrElf & \anElf & & $\star\star$ & & $\star\star$ & $\star\star$ & & & \\ %& 5
			\hline
			\anNrDreiundzwanzig & \anDreiundzwanzig & & & & & & $\star$ & $\star\star$ & \\%& 7
			\hline
			\anNrEins & \anEins & & & & & $\star$ & $\star$ & $\star\star$ & \\%& 9
			\hline
			\anNrVier & \anVier  & & & & & & & $\star\star$ & \\%& 14
			\hline
			\anNrSechs & \anSechs & $\star$ & & & & & & $\star\star$ & \\%& 21
			\hline
		\end{tabular}
		
	\vspace*{1ex}
	\caption{Zuordnung der zur Abfrage genutzten Methoden und der damit abgefragten Anforderungen. Ein $\star\star$ zeigt eine starke Verbindung der Methode mit der Anforderung. Ein $\star$ zeigt eine schwache Verbindung mit der Anforderung. Je mehr Sterne in einer Zeile sind, desto gründlicher wurde die Anforderung abgefragt. Aussagen aus Abbildung \ref{sec:Evaluation:eUCRITEDokumentation:Auswertung:Bild}.
	}
	\label{tab:UebersichtDerMethoden}
\end{table*}
}{
%short version \longVersionURL 
\begin{table*}
	\centering
	    \resizebox{2.0\columnwidth}{!}{%
		\begin{tabular}{|l|L{6.5cm}|c|c|c|c|c|c|c|c|}
			\hline
			$\#$ & {
			\diagbox[width=6.5cm,innerleftsep=.5cm,innerrightsep=0pt]
			{\textbf{Anforderung}}
			{ 
			%\begin{sideways} 
			    \textbf{Methode}  
			%\end{sideways}
			}
			} & {
%			\begin{sideways} \raggedleft  
			    Q\&A Fragen  
%			\end{sideways}
			} & {
%			\begin{sideways} 
			    Aussage 1 %supported me in solving the task 
%			\end{sideways}
			} & {
%			\begin{sideways} 
			    Aussage 2 %Found my way esaily 
%			\end{sideways}
			} & {
%			\begin{sideways} 
			    Aussage 3 %by accessing, i found useful help  
%			\end{sideways}
			} & {
%			\begin{sideways} 
			    Aussage 4 %helpful expations
%			\end{sideways}
			} & {
%			\begin{sideways} 
			    Aussage 5 %helpful code  
%			\end{sideways}
			} & {
%			\begin{sideways} 
			    Eye-Tracking  
%			\end{sideways}
			} & {
%			\begin{sideways}
			    Benötigte Zeit
%			\end{sideways}
			} \\
			\hline
			\anNrZweiundzwanzig & \anZweiundzwanzig & & & & $\star$ & & & & $\star\star$ \\ %& 1
			\hline
			\anNrDrei & \anDrei & & $\star$ & $\star\star$ & $\star$ & & & & \\ %& 2
			\hline
			\anNrSieben & \anSieben & & & & & & & $\star\star$ & \\ %& 3 
			\hline
			\anNrNeun & \anNeun & & & & & & & $\star\star$ & \\ %& 4 
			\hline
			\anNrElf & \anElf & & $\star\star$ & & $\star\star$ & $\star\star$ & & & \\ %& 5
			\hline
			\anNrDreiundzwanzig & \anDreiundzwanzig & & & & & & $\star$ & $\star\star$ & \\%& 7
			\hline
			\anNrEins & \anEins & & & & & $\star$ & $\star$ & $\star\star$ & \\%& 9
			\hline
			\anNrVier & \anVier  & & & & & & & $\star\star$ & \\%& 14
			\hline
			\anNrSechs & \anSechs & $\star$ & & & & & & $\star\star$ & \\%& 21
			\hline
		\end{tabular}
		}
	\vspace*{1ex}
	\caption{Zuordnung der zur Abfrage genutzten Methoden und der damit abgefragten Anforderungen. Ein $\star\star$ zeigt eine starke Verbindung der Methode mit der Anforderung. Ein $\star$ zeigt eine schwache Verbindung mit der Anforderung. Je mehr Sterne in einer Zeile sind, desto gründlicher wurde die Anforderung abgefragt. Aussagen aus Abbildung \ref{sec:Evaluation:eUCRITEDokumentation:Auswertung:Bild}.
	}
	\label{tab:UebersichtDerMethoden}
\end{table*}
}
Ziel dieser Studie war es, die Benutzbarkeit anhand der in der Literatur gefundenen Anforderungen (siehe Abschnitt \ref{sec:anforderungen}) der prototypischen Dokumentation in Kombination mit der \zeierAPI-API zu evaluieren.
Zur Vergleichbarkeit des Usability Scores nach Acar et al. \cite{acar_comparing_2017} wurde zusätzlich die Tink-API\footnote{Tink-API: \url{https://github.com/google/tink} Abgerufen am 16.12.2019} von Google untersucht.
%Für eine andere Arbeit den Usability Score nach Acar et al. \cite{acar_comparing_2017} zu ermitteln, wurde zusätzlich die Tink-API\footnote{Tink-API: \url{https://github.com/google/tink} Abgerufen am 16.12.2019} von Google untersucht. 
Die Tink-API ist eine plattformübergreifende Open-Source-Bibliothek, welche kryptografische APIs zur Verfügung stellt. Sie wurde unter dem Aspekt ausgewählt, dass sie gleichermaßen den Anspruch hat gut benutzbar zu sein und sich so für den Usability Score Vergleich eignet. In dieser Studie spielt die Tink-API eine untergeordnete Rolle, da ihre Dokumentation auf Github realisiert und ein expliziter Vergleich mit unserem Dokumentationssysten Unschärfen beinhaltet (siehe Abschnitt \ref{sec:Limitierung}).  
%(aneben wurde auch die \zeierAPI-API selbst hinsichtlich der Benutzbarkeit untersucht).
Für diese Studie sollten Probanden (Entwickler) einen fiktiven Chat-Client programmieren. Dazu bekamen sie zunächst ein Code-Grundgerüst zur Verfügung gestellt, in dem sie die Funktionen zur Generierung des Schlüsselmaterials, der Ver- und Entschlüsselung sowie der Erzeugung und Verifizierung von digitalen Signaturen implementieren sollten. Bei Unklarheiten wurden sie aufgefordert, sich in der entsprechenden Online-Dokumentation zu informieren. Dabei wurden sie über ein Eye-Tracking-System\footnote{Tobii Pro Studio \url{https://www.tobiipro.com/de/produkte/tobii-pro-studio/} Abgerufen am 16.12.2019} beobachtet.
Zum Vergleich sollte die gleiche Aufgabe von den Teilnehmern ebenfalls mit der Tink-API durchgeführt werden. Die Reihenfolge der APIs wurde anhand der Teilnehmernummer alternierend festgelegt. In Pre-Tests wurde eine Zeit von ca. 45 Minuten pro API ermittelt. Sollte der Teilnehmer für die Lösung der ersten Aufgabe länger als eine Stunde benötigt haben, so ist die zweite Aufgabe entfallen.

Die Aufgabenstellung und Fragen wurden in englischer Sprache gestellt. In einem Fragebogen wurden zu Beginn einige Informationen über die Teilnehmer abgefragt und später eine Bewertung der jeweiligen APIs erfasst. Hierzu wurden neben den von uns formulierten Fragen der von Acar et al. \cite{acar_comparing_2017} beschriebene API Usability Score verwendet.
\ifthenelse{\boolean{long}}{
%long version
Die genau gestellten Fragen und der Ablauf sind im Anhang \ref{sec:Anhang:Evaluation:eUCRITEvsTink} zu finden. 
}{
%short version \longVersionURL 
Die genau gestellten Fragen und ein detaillierter Ablauf sind in einer Langversion\footnote{Langversion: \longVersionURL} der Arbeit zu finden.
}
%Um neben dem Fragebogen von Acar et al. die Studie nicht zu lang werden zu lassen, wurden nur einige der in Abschnitt \ref{sec:anforderungen} beschriebenen Anforderungen evaluiert. 
Tabelle \ref{tab:UebersichtDerMethoden} zeigt 
die Zuordnung der zu untersuchenden Anforderungen zu den im Rahmen der Studie
eingesetzten Methoden (die konkreten Q\&A Fragen sowie Aussagen $1$--$5$ sind zusammen mit 
den Ergebnissen in den Abbildungen
\ref{img:Anhang:Evaluation:eUCRITEvsTink:initFragenJaNein} und
\ref{sec:Evaluation:eUCRITEDokumentation:Auswertung:Bild}  zu finden).
 Die Aussagen wurden von den Probanden in der Rohrmann-Skala \cite{rohrmann_empirische_1978} bewertet.
%Aussage 1 soll feststellen, wie hilfreich die Dokumentation generell bei der Bearbeitung der Programmieraufgabe ist. Durch Aussage 2 soll die Anforderung \anNrDrei\ nach einer \emph{guten Struktur} überprüft werden. Die weiteren Aussagen 3 - 5 stammen aus der Arbeit von Acar et al. \cite{acar_comparing_2017} zur Bestimmung des API Usability Scores. Wobei Aussage 3 der Anforderung \anNrZweiundzwanzig\ nahe kommt, schnell eine brauchbare Hilfe bzw. Einstieg zu finden. Durch die Beantworteung der Frage Nr. 6 wurden die Probanden nach einer Gesamteinschätzung im Schulnotensystem ($1$ = Sehr gut, ..., $6$ = ungenügend) gebeten.
%Die Teilnehmer wurden aus einem akademischen und industriellen Umfeld akquiriert. 
Die Teilnahme erfolgte freiwillig und unter den geltenden datenschutzrechtlichen Gegebenheiten. Als Aufwandsentschädigung bekamen die Teilnehmer im Anschluss einen $20$ Euro Amazon-Gutschein ausgehändigt.

\section{Ergebnisse}
\label{sec:Ergebnisse}
%\begin{figure}[t]
  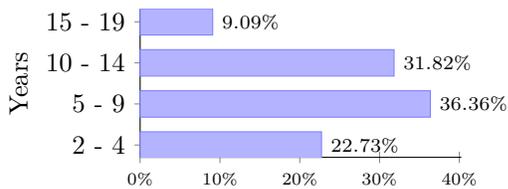
\begin{figure}
%\begin{minipage}{.4\textwidth}
\centering
\tikzstyle{every node}=[draw=none]
\pgfplotstableread[col sep=comma, header=true]{
age,man,woman
15 - 19,2,0
10 - 14,7,0
5 - 9,8,0
2 - 4,5,0
}\loadedtable
\begin{tikzpicture}[scale=1.0]
\begin{axis}[ 
%at={(popaxis.north west)},anchor=north east, %xshift=-3cm,
anchor=north west, %xshift=-3cm,
scale only axis,
    xbar = 0,
    xmin = 0,
    xmax = 35,
    width=0.24*\textwidth,
    height= 0.09*\textheight,
    y dir=reverse,
    nodes near coords = {\pgfmathprintnumber\pgfplotspointmeta\%},
    every node near coord/.append style={rotate = 0, anchor = west, font=\scriptsize, color=black},
    xticklabel= {\pgfmathprintnumber\tick\%},
    axis x line=left,
    axis y line*=left,
%    label style={font=\tiny}, %SChriftgröße Achsenbezeichnung
    xticklabel style = {font=\tiny}, %Schiftgröße Einheitenbezeichnung
    ytick = data,
    yticklabels from table = {\loadedtable}{age},
    ylabel=Years,
    ytick align=center,
    ytick pos=left,
    enlarge x limits = {value=0.15,upper},
    axis line style={-}
]
\addplot[blue!50,fill=blue!30] table[y expr =\coordindex, x expr={\thisrow{man}/22*100}] \loadedtable; %total pop = 31395 
\end{axis}
\end{tikzpicture}
\caption{Programmiererfahrung der Probanden}
	\label{img:Anhang:Evaluation:eUCRITEvsTink:Programmiererfahrung}
%\end{minipage}
\end{figure}
  %\begin{minipage}{.6\textwidth}
\begin{figure}
\centering
\tikzstyle{every node}=[draw=none]
\pgfplotstableread[col sep=comma,header=true]{
Frage, Ja, Nein
Have you ever worked with a cryprographic API?,9,13
Did you ever ask a question at a Q\&A-Platform?,12,9
Did you ever answer a question at a Q\&A-Platform?,6,16
}\data

\pgfplotsset{
percentage plot/.style={
    point meta=explicit,
    xticklabel=\pgfmathprintnumber{\tick}\,$\%$,
    xmin=0,
    xmax=100,
    enlarge x limits={upper,value=0},
visualization depends on={x \as \originalvalue}
},
}

\begin{tikzpicture}

\begin{axis}[ 
%at={(popaxis.north west)},anchor=north east, %xshift=-3cm,
anchor=north west, %xshift=-3cm,
legend style={at={(1.05,0.85)} },
reverse legend,
scale only axis,
    xbar = 0,
    xmin = 0,
    width=0.24*\textwidth,
    height= 0.185*\textheight, %<-Abstand zwischen den Balken
    y dir=reverse,
    nodes near coords = {\pgfmathprintnumber\pgfplotspointmeta\%},
    every node near coord/.append style={rotate = 0, anchor = west, font=\scriptsize, color=black},
    xticklabel= {\pgfmathprintnumber\tick\%},
    axis x line=left,
    axis y line*=left,
    ytick = data,
    yticklabels from table = {\data}{Frage},
    ytick align=center,
    ytick pos=left,
	yticklabel style={text width=3.5cm,font=\small}, %<- Text umbruch bei y lable
	xticklabel style = {font=\tiny}, %Schiftgröße Einheitenbezeichnung
    enlarge x limits = {value=0.15,upper},
    axis line style={-}
]

%\addplot[blue!50,fill=blue!30] table[y expr =\coordindex, x expr={\thisrow{man}/22*100}] \loadedtable;
\addplot table[y expr =\coordindex, x expr={\thisrow{Ja}/(\thisrow{Ja}+\thisrow{Nein})*100}] \data;
\addplot table[y expr =\coordindex, x expr={\thisrow{Nein}/(\thisrow{Ja}+\thisrow{Nein})*100}] \data;
\legend{Yes,No}
\end{axis}
\end{tikzpicture}
%\caption{Ergebnisse der Ja/Nein-Fragen aus dem Fragebogen.}
\caption{Ergebnisse der Ja/Nein-Fragen.}
	\label{img:Anhang:Evaluation:eUCRITEvsTink:initFragenJaNein}
 \end{figure}
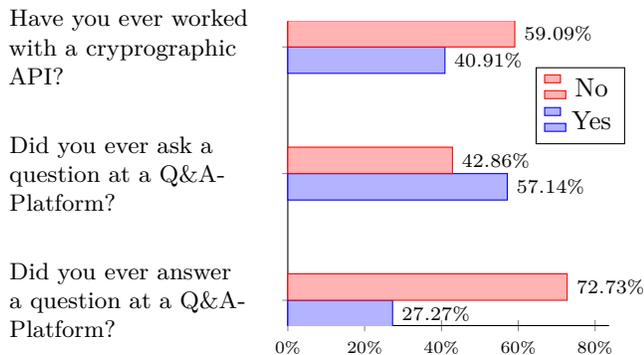
%\end{minipage}
%\end{figure}
\ifthenelse{\boolean{long}}{
%long version
\begin{figure*}[t]
\centering
  \includegraphics[width=0.8\textwidth]{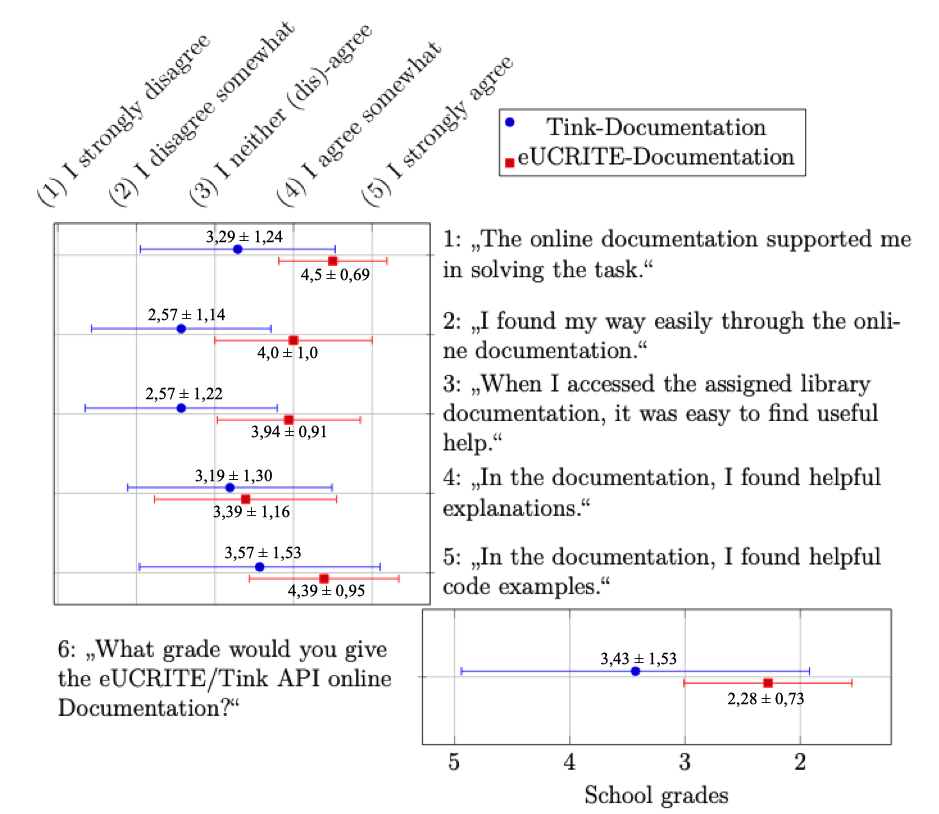}
  \caption{Der arithmetische Mittelwert und die Standardabweichung (als Fehlerbalken) der Aussagen 1 -- 5 und Frage 6 in Schulnoten.}
  \label{sec:Evaluation:eUCRITEDokumentation:Auswertung:Bild}
\end{figure*}
}{
%short version 
\begin{figure}[t]
\centering
  \includegraphics[width=0.5\textwidth]{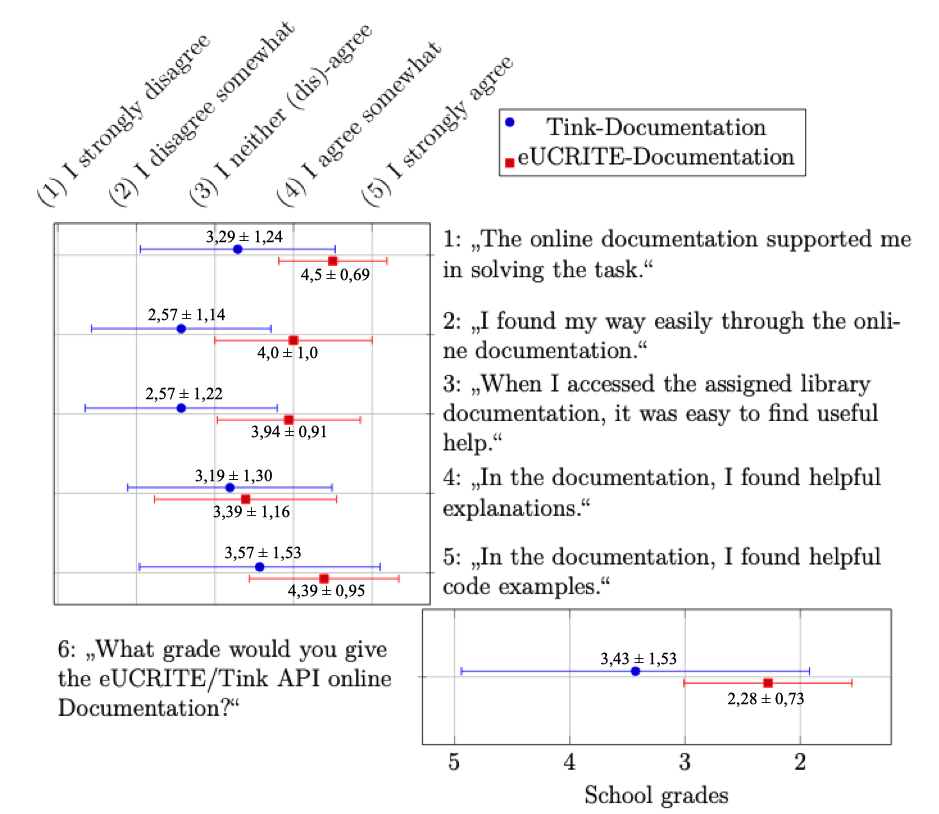}
  \caption{Der arithmetische Mittelwert und die Standardabweichung (als Fehlerbalken) der Aussagen 1 -- 5 und Frage 6 in Schulnoten.}
  \label{sec:Evaluation:eUCRITEDokumentation:Auswertung:Bild}
\end{figure}
}
Insgesamt haben $22$ Probanden an der Studie teilgenommen. $6$ davon aus der Industrie. Bei den ersten acht Probanden wurde das Eye-Tracking-System nicht richtig konfiguriert, sodass deren Aufzeichnungen keine effiziente Auswertung zuließ. Mit dem Eye-Tracking-System haben $13$ Probanden die Tink- und $12$ die \zeierAPI-Programmieraufgabe  bearbeitet.  
Die Probanden haben durchschnittlich $30$ Minuten für das Lösen der \zeierAPI- und $43$ Minuten für das Lösen der Tink-Programmierauf\-gabe benötigt. Die Studiendauer lag im Durchschnitt bei $93$ Minuten. 
Die durchschnittliche Programmiererfahrung der Probanden liegt bei $8$ Jahren (Median: $7$ Jahre). Die Verteilung ist in  Abbildung \ref{img:Anhang:Evaluation:eUCRITEvsTink:Programmiererfahrung} zu sehen. 
Für $13$ Probanden ($59,09\%$) stellt diese Studie zum ersten Mal ein Berührungspunkt mit einer Kryptographie-API dar. Die Frage, ob sie schon mal eine Frage auf einer Q\&A-Plattform gestellt haben, beantworteten $12$ Probanden ($57,14\%$) mit ja. 
$6$ Probanden ($27,27\%$) haben schon ein Mal auf eine Frage in einer Q\&A-Plattform geantwortet.
Abbildung \ref{img:Anhang:Evaluation:eUCRITEvsTink:initFragenJaNein} zeigt die Ergebnisse als Balkendiagramm. 

\begin{figure}
 \centering
 \includegraphics[width=0.49\textwidth]{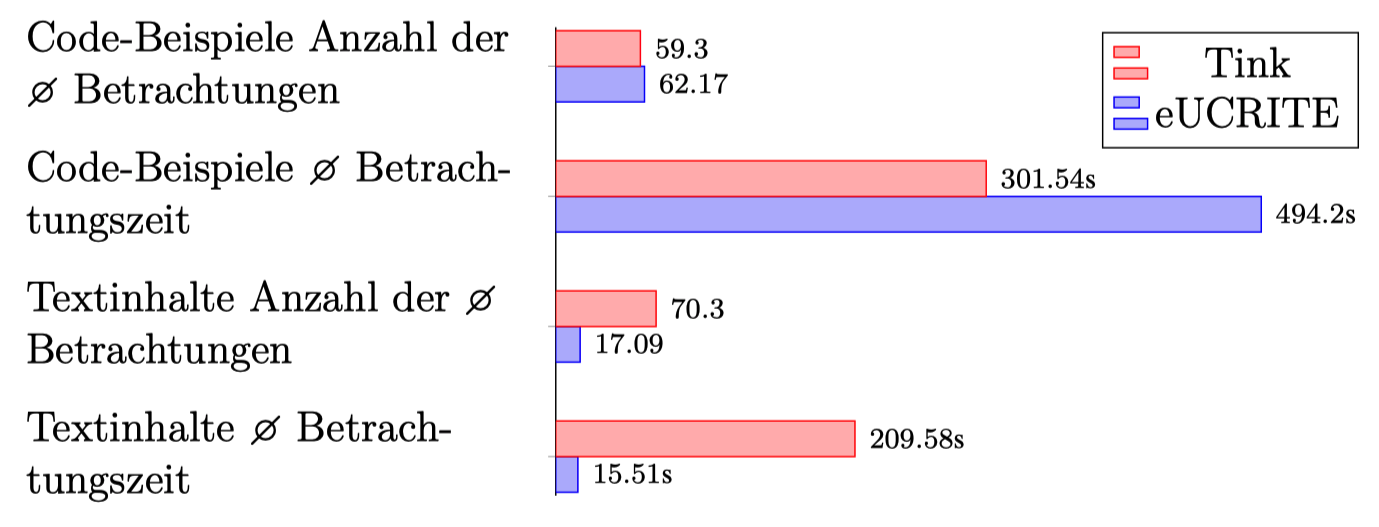}
 \caption{AOI-Gruppen der Tink- und \zeierAPI-Dokumentation im Vergleich.}
 \label{img:Evaluation:aoi}
\end{figure}

In dieser Auswertung werden ausschließlich die Fragen zur prototypischen Dokumentation mit der \zeierAPI-API betrachtet. 
Die von den Probanden bewerteten Aussagen sind in Abbildung \ref{sec:Evaluation:eUCRITEDokumentation:Auswertung:Bild} mit den arithmetischen Mittelwerten und den Standardabweichungen abgebildet.
Bei Betrachtung der Abbildung \ref{sec:Evaluation:eUCRITEDokumentation:Auswertung:Bild} kann für die \zeierAPI-API Dokumentation gefolgert werden, dass die Beispiele (Aussage 5) in der Dokumentation geholfen haben, die gestellte Aufgabe zu lösen (Aussage 1). Das Ergebnis der Aussage 4 kann daraus resultieren, dass viele Probanden eventuell eine  opportunistische Arbeitsweise (siehe Clark \cite{clarke_what_2007}) gewählt haben und dadurch die Textinhalte nicht gelesen haben. 
\ifthenelse{\boolean{long}}{
%long version
\begin{figure}
 \centering
 \includegraphics[trim = 6cm 8cm 12cm 2cm,clip,width=0.5\textwidth]{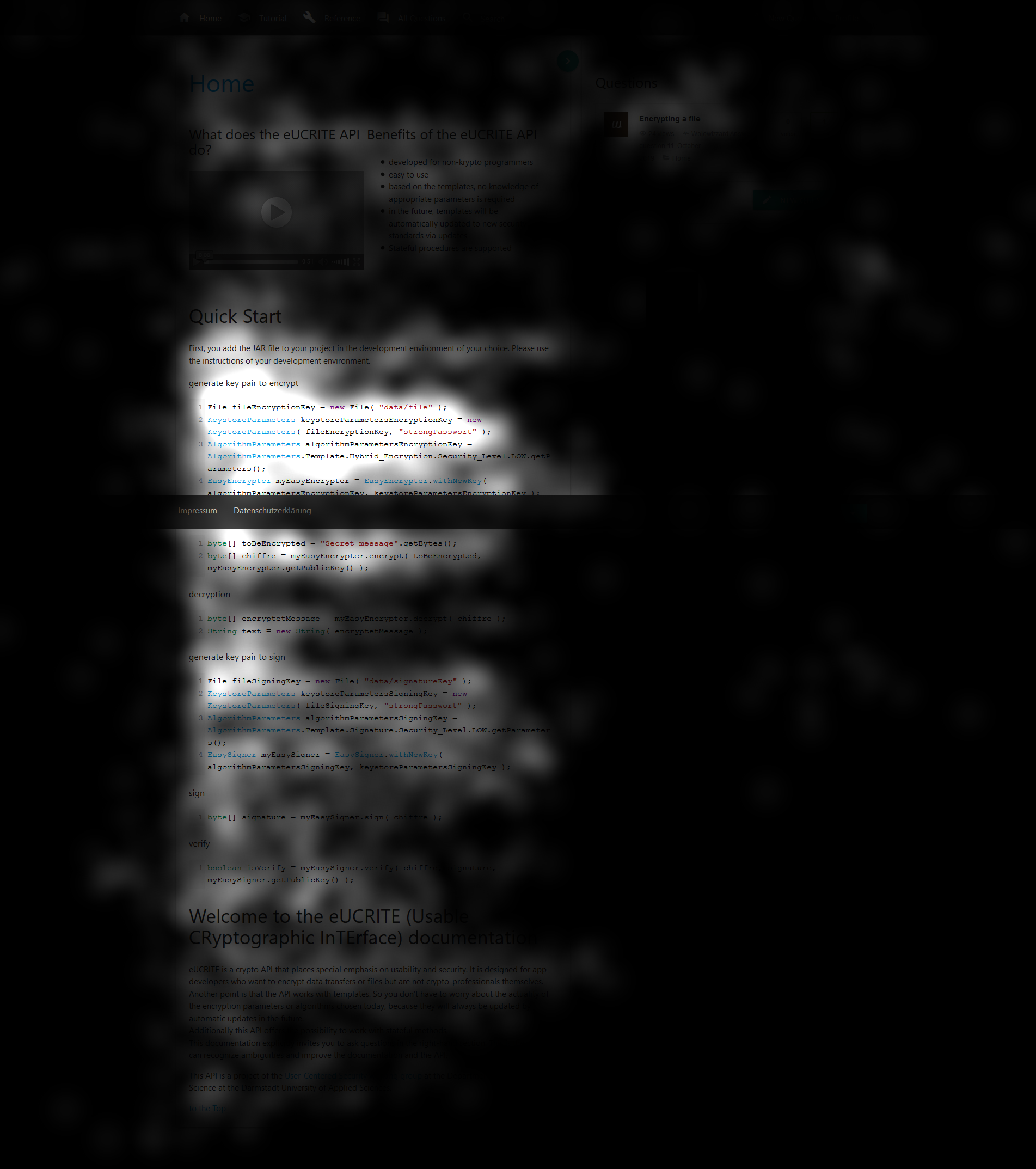}
 \caption{Negativ Headmap der \zeierAPI-API Dokumentation Startseite.}
 \label{fig:Evaluation:eUCRITEDokumentation:NegativHeatmapSatrtseite}
\end{figure}
Diese These wird in der vom Eye-Tracking-System generierten negativ Heatmap der Startseite in Abbildung \ref{fig:Evaluation:eUCRITEDokumentation:NegativHeatmapSatrtseite} unterstützt.
}{
%short version \longVersionURL 
Diese These wird in der vom Eye-Tracking-System generierten negativ Heatmap\footnote{Die negativ Heatmap ist in der Langversion der Arbeit zu finden: \longVersionURL} der Startseite unterstützt.
} 
Dort ist zu erkennen, dass die Probanden oft das Code-Beispiel zum Generieren der Schlüssel zur Verschlüsselung betrachtet haben. 
Um dies mit dem Eye-Tracking-System zu evaluieren, wurden alle Code-Beispiele und Textinhalte in Area of Interests (AOI) eingeteilt und gruppiert. 
Allgemein lässt sich zu den AOI-Gruppen feststellen, dass die Code-Beispiele $\diameter 62,17$ Mal und $\diameter 494,2$ Sekunden lang von Probanden studiert wurden. Textinhalte und Beschreibungen wurden pro Proband mit $\diameter 17,09$ Ansichten und $\diameter 15,51$ Sekunden Aufmerksamkeit benutzt. Damit wird den Textinhalten gerade einmal $3\%$ der Betrachtungszeit von Code-Beispielen gewidmet. Dies bestätigt die Vermutung, dass viele Probanden eine opportunistische Arbietsweise (siehe Clarke \cite{clarke_what_2007}) gewählt haben.
Zum Vergleich wurde bei der Tink-Dokumentation festgestellt, dass die Code-Beispiele $\diameter 59,3$ Mal und $\diameter 301,54$ Sekunden lang studiert wurden. Die Textinhalte wurden $\diameter 70,3$ Mal und $\diameter 209,58$ Sekunden betrachtet. 
Diese Zahlen lassen darauf schließen, dass die Probanden eine pragmatische Arbeitsweise angenommen haben, da sie vergleichsweise kürzer auf die Code-Beispiele aber öfter und länger auf die Textinhalte schauten. %, als dies bei dem opportunistischen Verhalten bei Verwendung der \zeierAPI-API der Fall war. 
Der Vergleich lässt sich in Abbildung \ref{img:Evaluation:aoi} ablesen. 
Viele Probanden haben in der \zeierAPI-API-Doku\-mentation ausschließlich die Startseite genutzt. Proband Nr. $5$ hat auf die Frage, was bei der Online-Dokumentation weiter verbessert werden kann folgendes angegeben: \textit{"`Bessere Struktur. Das Tutorial habe ich erst ganz zum Schluss gefunden"'}. Proband Nr.~$2$ schrieb: \textit{"`Auf [der] Startseite [sind] wenig erklärende Elemente zum Code"'}. Er hat dem Tutorial, in dem mehr erklärende Elemente zum Code vorhanden sind, beziehungsweise dem Header-Menü, keine Auf\-merk\-sam\-keit geschenkt. Die Klick-Zahlen der Eye-Tracking-Aufzeichnungen unterstützen die These, dass viele Probanden ausschließlich die Startseite genutzt haben. Von den $12$ Probanden haben ausschließlich $5$ die Seite "`Tutorial"' aufgerufen. Wir vermuten, dass die Probanden das Header-Menü nicht als Menü erkannt haben oder die Informationen auf der Startseite völlig ausreichend waren.
\ifthenelse{\boolean{blindForReview}}{
%Anonym
Dies untermauert das in der Arbeit von Huesmann et al. \cite{huesmann_eigenschaften_2019} beschriebene Interaktionsmuster, 
}{
%Autorennamen
Dies untermauert das in unserer vorherigen Arbeit \cite{huesmann_eigenschaften_2019} beschriebene Interaktionsmuster, 
} 
nach dem sich Probanden bevorzugt über den Schnelleinstieg (was dem Tutorial in \cite{huesmann_eigenschaften_2019} entspricht) in das für sie neue Themengebiet der \zeierAPI-API eingearbeitet haben. 

Der Absprungrate in die klassische Referenz zufolge haben drei Probanden bemerkt, dass die blauen Klassen-Namen in den Code-Beispielen Querverweise (Hyper-Links) in die Referenz darstellen (Anforderung \anNrSieben). Die anderen $12$ Teilnehmer haben die blaue Schrift vermutlich als Syntax-Highlighting interpretiert.% Über einen auffälligeren "`Mouse-Over"' Effekt könnte dies verdeutlichter werden.

Das Eye-Tracking-System hatte Schwierigkeiten den Bereich des Header-Menüs zu verarbeiten. Weil das Header-Menü über CSS am Browserfenster fixiert ist, folgt es dem Browserfenster beim Scrollen. Dies vollzieht das Eye-Tracking-System nicht nach und ordnet die Blicke auf das Header-Menü und dem Footer Stellen auf der Webseite zu.
Daher sind die Eye-Tracking-Zahlen im Header-Menü mit Vorsicht zu interpretieren und es wurden ausschließlich die Blicke von Probanden im Header-Menü gezählt, wenn die Webseite sich im Urzustand befand und nicht gescrollt wurde. Dementsprechend ist davon auszugehen, dass die tatsächlichen Zahlen höher liegen als in Tabelle \ref{sec:Evaluation:eUCRITEvsTink:Auswertung:AOIMenu} aufgelistet.
%
%\newcolumntype{L}[1]{>{\raggedright\arraybackslash}p{#1}} % linksbündig mit Breitenangabe
%\newcolumntype{C}[1]{>{\centering\arraybackslash}p{#1}} % zentriert mit Breitenangabe
%\newcolumntype{R}[1]{>{\raggedleft\arraybackslash}p{#1}} % rechtsbündig mit Breitenangabe
\begin{table}
\centering
\ifthenelse{\boolean{long}}{
%long version
\begin{tabular}{|L{1.5cm}|C{1.8cm}|c|C{2cm}|}
\hline
Menü-Punkt & Anzahl Probanden & $\diameter$ Fokus & $\diameter$ Betracht\-ungszeit\\ 
\hline
Tutorial & $4$ & $1$ & $0,24$ Sek. \\ 
\hline
Referenz & $5$ & $2,4$ & $0,7$ Sek.\\
\hline
Suchfeld & $3$ & $1$ & $0,24$ Sek.\\
\hline 
\end{tabular}
}{
%short version \longVersionURL 
\resizebox{0.8\columnwidth}{!}{%
\begin{tabular}{|L{1.5cm}|C{1.8cm}|c|C{2cm}|}
\hline
Menü-Punkt & Anzahl Probanden & $\diameter$ Fokus & $\diameter$ Betracht\-ungszeit\\ 
\hline
Tutorial & $4$ & $1$ & $0,24$ Sek. \\ 
\hline
Referenz & $5$ & $2,4$ & $0,7$ Sek.\\
\hline
Suchfeld & $3$ & $1$ & $0,24$ Sek.\\
\hline 
\end{tabular}
}
}
\caption{AOIs des Header-Menüs der Startseite}
\label{sec:Evaluation:eUCRITEvsTink:Auswertung:AOIMenu}
\end{table}
% Erstmal mit positiven Ergebnissen anfangen. 

Die Aussage von Proband Nr. $13$: \textit{"`Besonders gefallen hat mir die übersichtliche Doku und die einfachen Codebeispiele"'} unterstützt die Bewertungen der Aussagen 1 und 5. Da zu jeder Methode, Klasse und Funktion ein Beispiel vorhanden ist, kann daraus mit der Aussage des Probanden gefolgert werden, dass Anforderung \anNrEins\ \emph{\anEins} erfüllt wurde.  
Anforderung \anNrSechs\ \emph{\anSechs} wurde umgesetzt. Laut der Antworten in Abbildung~\ref{img:Anhang:Evaluation:eUCRITEvsTink:initFragenJaNein} ist die Hälfte der Probanden mit der Funktionsweise vertraut. 11 Probanden haben die Q\&A-Fragen $\diameter 28$ Mal und $\diameter 19,01$ Sekunden betrachten. 2 Probanden haben aktiv mit ihnen interagiert.
Da viele Probanden anscheinend genügend Informationen zum Erfüllen der Programmieraufgabe mit der für sie unbekannten \zeierAPI-API auf der Startseite gefunden und $\diameter 30$ Minuten zum Lösen der Programmieraufgabe benötigt haben, kann daraus geschlossen werden, dass Anforderung \anNrZweiundzwanzig\ \emph{\anZweiundzwanzig} erfüllt ist.
Anforderung \anNrDreiundzwanzig\ \emph{\anDreiundzwanzig} unterstützt die opportunistische Arbeitsweise. Da die Probanden dies bevorzugt in der \zeierAPI-API-Dokumentation angewendet haben (siehe Schlussfolgerung weiter oben), lässt dies den Schluss zu, dass Anforderung \anNrDreiundzwanzig\ erfüllt ist.

%Abbildung \ref{img:Evaluation:aoi} zeigt, dass sich die Arbeitsweise der Probanden zwischen der \zeierAPI- und Tink-API-Doku\-men\-ta\-tion unterscheiden. In der \zeierAPI-API-Doku\-men\-tation haben die Probanden eine opportunistischer Arbeitsweise angenommen. In der Tink-API-Dokumenta\-tion haben die Probanden eine pragmatische Arbeitsweise angenommen. Dies lässt sich aus der kurzen Betrachtungszeit der Code-Beispiele und der gestiegenen Anzahl und Betrachtungszeit der Texinhalte in Abbildung \ref{img:Evaluation:aoi} ableiten.
% (Die Benutzbarkeit der \zeierAPI-API selbst wurde mit einem Usability Score nach Acar et al. \cite{acar_comparing_2017} von $70,5$ bewertet. Im Vergleich erreichte die Tink-API einen Score von $48,23$.)

\section{Limitierung}\label{sec:Limitierung}

Neben den bereits benannten Limitierungen, weist unsere Studie und die Ergebnisse die folgenden weiteren Limitierungen auf: 
%Eine Dokumentation 
Eine Aussage, ob das Dokumentationssystem verhindert, dass Programmiere unsicheren Code schreiben, kann nicht getroffen werden, da die Tink- und \zeierAPI-API über ihr Design bereits versuchen dies zu verhindern.
Aus Zeit- und Ressourcengründen haben wir mit unserer Nutzerstudie zwei unterschiedliche Ziele verfolgt. Neben der Evaluierung des in dieser Arbeit vorgestellten Dokumentationssytems wurde zusätzlich die Benutzbarkeit der \zeierAPI-API selbst im Vergleich zur Tink-API untersucht. (Die Benutzbarkeit der \zeierAPI-API selbst wurde mit einem Usability Score nach Acar et al. \cite{acar_comparing_2017} von $70,5$ bewertet. Im Vergleich erreichte die Tink-API einen Score von $48,23$.)

Da die Tink-API in der Github-Umgebung dokumentiert ist, bringt ein direkter Vergleich mit dem prototypischen Dokumentationssystem eine Unschärfe mit sich.
Allerdings stehen die Erkenntnisse der Evaluation der prototypischen Umsetzung des Dokumentationssystem für sich. 
Die Teilnehmer wurden aus dem lokalen Umfeld der Autoren gewonnen, somit könnten diese wohlwollend gehandelt haben und spiegeln keine repräsentative Gruppe wider.

\section{Zusammenfassung und Ausblick}
\label{sec:Zusammenfassung}
Uns ist kein Dokumentationssystem bekannt, welches alle in der Literatur aufgeführten Anforderungen in seiner Gesamtheit umsetzt. In dieser Arbeit wurden diese Anforderungen aus der Literatur extrahiert und in Teilen prototypisch umgesetzt. Hierbei wurden die Anforderungen zweigeteilt. Zum einen die Anforderungen, welche von den bereitgestellten Informationen erfüllt werden sollten. 
\ifthenelse{\boolean{long}}{
%long version
Aus diesen wurde eine Checkliste (siehe Anhang \ref{sec:Anhang:checkliste}) zur Orientierung für Dokumentationsersteller erarbeitet.
}{
%short version \longVersionURL 
Aus diesen wurde eine Checkliste zur Orientierung für Dokumentationsersteller erarbeitet.
} 
 Zum anderen in Anforderungen, welche ein optimales Dokumentationssystem erfüllen sollte.
\ifthenelse{\boolean{long}}{
%long version
Aus diesen Anforderungen wurde ein Konzept (siehe Anhang \ref{sec:Anhang:konzept}) eines optimierten Dokumentationssystems entworfen. 
}{
%short version \longVersionURL 
Aus diesen Anforderungen wurde ein Konzept eines optimierten Dokumentationssystems entworfen. 
} 
Auf der Grundlage dieses Konzepts wurde eine prototypische Implementierung  eines optimierten Dokumentationssystems erstellt. Im Anschluss wurde in diesem Dokumentationssystem die \zeierAPI-API-Dokumentation exemplarisch umgesetzt.% und parallel zu dieser Studie hinsichtlich der Benutzbarkeit evaluiert. 

Diese Arbeit kommt zu dem Ergebnis, dass einige der gestellten Anforderungen im Prototypen erfüllt werden konnten. Zum Beispiel die Anforderung \anNrEins\ \emph{\anEins}, \anNrSechs\ \emph{\anSechs}, \anNrZweiundzwanzig\ \emph{\anZweiundzwanzig} und \anNrDreiundzwanzig\ \emph{\anDreiundzwanzig}.
Bei weiteren besteht jedoch noch Verbesserungsbedarf. Beispielsweise könnte durch eine Umgestaltung des Header-Menüs vermutlich eine Verbesserung der Anforderungen \anNrDrei\ \emph{\anDrei}, \anNrVier\ \emph{\anVier}, \anNrNeun\ \emph{\anNeun} und \anNrElf\ \emph{\anElf} erzielt werden. Dies soll durch weitere Studien untersucht werden.
Es wurde festgestellt, dass die prototypische Implementierung der optimierten Dokumentation die opportunistische Arbeitsweise von Entwicklern nach Clarke \cite{clarke_what_2007} unterstützt. In weiteren Studien soll untersucht werden, ob die Dokumentation für pragmatisch und systematisch arbeitende Entwickler ebenso geeignet ist. 
Insgesamt wurde die prototypische Implementierung von den Teilnehmern der Studie mit der Schulnote "`2--"' bewertet. Zum Vergleich wurde  die via Github gehostete Tink-API-Dokumentation mit einer "`3--"' bewertet.

In einem nächsten Schritt sollen die nicht betrachteten Anforderungen aus der Literatur in den Prototypen integriert und erneut in Gänze evaluiert werden.

Die Erkenntnisse aus dieser Arbeit sind nicht auf kryptografische APIs beschränkt. Allerdings sind die Auswirkungen von Programmierfehlern bei kryptografischen APIs aus unserer Sicht vergleichsweise gravierender \cite{wijayarathna_generic_2017}. Deswegen wurde im Kern dieser Arbeit eine kryptografische API betrachtet.

Eine Dokumentation ist nur so gut, wie sie ihr Ersteller einrichtet und pflegt. Ein Dokumentationssystem sollte nicht ausschließlich die Benutzbarkeit aus der Sicht des Softwareentwicklers betrachten, sondern zusätzlich dem Dokumentationsersteller gute Werkzeuge bieten, um die Dokumentation leicht und teilautomatisiert zu erstellen und zu pflegen. Dieser Aspekt wurde in dieser Arbeit nicht betrachtet, da das Augenmerk dieser Studie darauf gelegt wurde, welches Dokumentationssystem Softwareentwickler am besten unterstützt. In zukünftigen Arbeiten sollte untersucht werden, wie ein API-Dokumentationsersteller unterstützt werden kann, Inhalte einfach und benutzerfreundlich zu integrieren.

%Eine weitere offene Frage ist, wer bezahlt und pflegt die API-Dokumentation? Können API-Entwickler verpflichten werden, ihre API-Dokumentation lebenslang zu pflegen? Selbst wenn die API nicht mehr weiterentwickelt wird, sollten Informationen in der Dokumentation weiterhin aktualisiert werden. Wenn zum Beispiel ein benutztes Verschlüsselungsverfahren gebrochen wird, sollte in der Dokumentation eine Warnung eingepflegt werden. Alternativ könnte es eine Stiftung geben, welche die $x$ wichtigsten API-Dokumentation\-en pflegt. Diese könne beispielsweise durch eine freiwillige Finanzierung von IT-Dienstleistern unterhalten werden. Es sollte im Interesse ihrer Kunden bzw. zur Prävention negativer Schlagzeilen in ihrem eigenen Interesse, sein kryptografische API-Dokumentationen zu pflegen.

\section{Förderhinweis}
\label{sec:thx}
Diese Arbeit  wurde im Rahmen der Innovationsförderung Hessen aus Mitteln der LOEWE – Landes-Offensive zur Entwicklung Wissenschaftlich-ökonomischer Exzellenz, Förderlinie 3: KMU-Verbundvorhaben (Projekt HA-Projekt-Nr.: 633/18-5) gefördert sowie vom Bundesministerium für Bildung und Forschung (BMBF) und vom Hessischen Ministerium für Wissenschaft und Kunst (HMWK) im Rahmen ihrer gemeinsamen Förderung für das Nationale Forschungszentrum für angewandte Cybersicherheit ATHENE unterstützt.
%-----------------------------------------------------------------------------------------------------------------
% Literatur
%-----------------------------------------------------------------------------------------------------------------
\bibliographystyle{syssec}
\bibliography{literatur} % Hier Ihre bib-File eintragen, oder unsere Vorlage ergänzen

\ifthenelse{\boolean{long}}{

%\appendix
\part{Appendix}

%************************************************
\section{Konzept der optimierten Dokumentation}\label{sec:Anhang:konzept}
%************************************************

Hier wird das Konzept der optimierten Dokumentation pro Seite erläutert. Prinzipiell ist jede inhaltliche Seite in einen rechten Bereich in dem Fragen und Antworten geschrieben werden können, wie in Anforderung \anNrSechs\ gewünscht, und einem mittlerem Bereich, in dem der eigentliche Inhalt der Seite präsentiert wird, aufgeteilt. Ausnahmen sind Seiten wie die Suche, alle Fragen, organisatorische Seiten die zur Pflege des User-Profils erforderlich sind und gesetzlich vorgeschriebene Seiten wie zum Beispiel das Impressum und die Datenschutzverordnung.
\begin{description}

	\item[Home:] In dem Fall der Startseite ist im mittleren Bereich ein sehr kurzes Video vorgesehen, welches schnell einen Überblick über die Funktionalität der API geben soll. Daneben sollen in kurzen Stichpunkten die Vorteile bzw. die Alleinstellungsmerkmale der API hervorgehoben werden. Darunter ein Schnellstart vorhanden sein, wie in der Anforderung \anNrDreiundzwanzig\ gewünscht. Im Anschluss ist Platz für ein Text über die API und deren Programmierer bzw. Community oder ähnliches.
	
	\item[Tutorial:] Dort soll es für den API-Einsteiger ein "`Schritt-für-Schritt"' Praxis-Beispiel geben. Die Einleitung besteht aus einem kurzen Video, welches das Ziel des Tutorials erklärt und einem stichpunktartigen Inhaltsverzeichnis. Dies soll der Anforderung \anNrElf\ der \emph{verschiedenen Niveaus} gerecht werden. In einem Tutorial lassen sich gut \emph{\anZwoelf} und die \emph{Fehlerbehandlung} ansprechen (Anforderung \anNrZwoelf\ \& \anNrAcht).
	
	\item[Referenz:] Unter dem Punkt verbirgt sich Anforderung \anNrVier\ einer \emph{klassischen Referenz}, in der jede Klasse, Methode und Funktion mit allen Parametern beschrieben ist. Dazu sollte für jedes Package eine Auflistung der beinhalten Klassen aufgeführt sein. Da eigentlich jede Funktion eine eigene Unterseite haben sollte, kann als Hilfe rechts neben der Liste die fünf am meisten besuchten Seiten bzw. Funktionen aufgelistet werden.
	\begin{description}
		\item[Klassen:] Eine Unterseite der Referenz ist eine Klas\-se. Dort gibt es zur leichteren Orientierung einen Brotkrumen-Pfad. Darunter werden die Funktionen der Klasse aufgeführt und eine kurze Aufgabenbeschreibung präsentiert. Anschließend sollten Beispiele aufgeführt werden. Hier bieten sich Beispiele mit \emph{Fehlerbehandlung} an (Anforderung \anNrAcht).
		\begin{description}
		\item[Funktionen:] Eine Funktion ist unter der jeweiligen Klassen-Seite angeordnet. Dort wird neben den Parametern und mindestens einem Beispiel, ein kurzer Text angezeigt, welche Aufgaben die Funktion hat. Funktionen oder Methoden welche die gleiche oder ähnliche Aufgaben, eventuell mit anderen Parametern erfüllen, sollten auf dieser Seite gemäß Anforderung \anNrAchtzehn\ aufgelistet sein. 
		\end{description}
	\end{description}
	
	\item[Alle Fragen:] Dies ist eine Übersichtsseite auf der alle Fragen aufgelistet werden. Explizit für Entwickler, die nach Fragen suchen oder anderen antworten möchten.

	\item[Suche:] Die Suche ist in dem Menü-Balken als Eingabefeld realisiert. Beim drücken der Eingabetaste wird es abgeschickt und das Suchergebnis auf der Seite mit der erweiterten Suche angezeigt. Auf dieser kann die Suche nochmal mit weiteren Suchparametern, wie in Anforderung \anNrNeun\ gewünscht, verfeinert werden.

	\item[Neue Frage stellen:] Dieser Menü-Punkt soll es dem Entwickler ermöglichen auf jeder Seite und zu jeder Zeit eine Frage stellen zu können. Damit dies beim Scrollen von längeren Seiten möglich ist, ist das Header-Menü oben am Browserfenster fixiert.

	\item[Login:] Dies ist der Menü-Punkt, welcher zur Profilverwaltung führt, wie zum Beispiel: Passwort ändern, Account löschen oder Community-Ranking. Dies wird in dieser Arbeit nicht weiter behandelt und offengelassen.

\end{description}
%\graffito{\textcolor{red}{Das ist eher schon Funktionales Konzept des Dokumentationssystems.}}
Pro inhaltliche Seite sollen Fragen und Antworten zu der entsprechenden Seite gestellt werden können. Diese Fragen und Antworten werden durch eindeutige Kategorien (zum Beispiel dem Brotkrumen-Pfad) der jeweiligen Seite zugeordnet. Klassen-, Funktionen- und Methoden-Namen werden in Code-Beispielen, Beschreibungen, Fragen und Antworten automatisch mit der passenden Seite in der Referenz verlinkt.

%************************************************
\section{Checkliste für Dokumentationsersteller}\label{sec:Anhang:checkliste}
%************************************************
Diese Checkliste soll dem Dokumentationsersteller als Gedankenstütze helfen. Die einzelnen Punkte sind absteigend der Priorität sortiert.

  \begin{todolist}
    \item \textbf{\anEins} \\ Ungefähr für jede Funktion ein Beispiel.
    \item \textbf{\anZwei}
    \item \textbf{\anDrei} \\ Übersichtlichen, leicht zu navigierenden Seitenaufbau, eine effiziente Navigation durch die API-Dokumentation, einfach und zeitsparend.
    \item \textbf{\anAcht} \\ Exception Handling wird beispielhaft erläutert.
    \item \textbf{\anZehn} \\ Konzeptionelle Informationen sollten mit der Beschreibung von Aufgaben, geeigneten Beispielen oder Nutzungsszenarien präsentiert werden.
    \item \textbf{\anElf} \\ Zum Beispiel ein Tutorial für den Einsteiger und eine Referenz oder spezielle Beispiele für Anwender mit mehr Hintergrundwissen.
    \item \textbf{\anZwoelf} \\ Typische Fehler oder Fallen in Negativbeispielen hervorheben.
    \item \textbf{\anDreizehn} \\ Wenn es in das Konzept der API passt.
    \item \textbf{\anVierzehn} \\ In regelmäßigen Abständen überprüfen.
    \item \textbf{\anFuenfzehn} \\ Ist wirklich alles dokumentiert?
    \item \textbf{\anAchtzehn} \\ Gibt es Methoden welche die gleichen oder ähnliche Aufgaben, eventuell mit anderen Parametern, übernehmen?
    \item \textbf{\anZwanzig} \\ Wenn Entwickler bei der Arbeit mit der API auf ein Problem stoßen und sich an die API-Dokumentation wenden, um Informationen zu finden, die das Problem lösen, kennen sie wahrscheinlich die Inhaltsdomäne ihres Problems.\\Bedeutet, das die Lösung zu einem Problem zur jeweiligen Klasse oder Funktion in der Referenz als Beispiel aufgezeigt werden sollte.
    \item \textbf{\anZweiundzwanzig} \\ Gibt es einen Schnelleinstieg um zeitnahe Erfolgserlebnisse zu gewährleisten?
    \item \textbf{\anVierundzwanzig} \\ Gibt es kritische Informationen? Werden diese an allen wichtigen Orten in der Dokumentation aufgegriffen?
    \item \textbf{\anFuenfundzwanzig} \\ Gibt es Einstellungen oder Optionen, welche die Performance beeinträchtigen?
  \end{todolist}

%************************************************
\section{Vergleichsstudie eUCRITE vs. Tink}\label{sec:Anhang:Evaluation:eUCRITEvsTink}
%************************************************
%TODO: kurz Ablauf beschreiben.\graffito{\textcolor{red}{fertig schreiben}}
Am Anfang durften die Probanden die "`Einwilligungserklärung zur Erhebung und Verarbeitung personenbezogener Daten für Forschungszwecke"' lesen und unterschreiben.
Anschließend wurde ihnen die Aufgabenstellung \ref{sec:Anhang:Evaluation:eUCRITEvsTink:aufgabe} und die Arbeitsumgebung erklärt. Zuerst sollten die Probanden den Fragebogen \ref{sec:Anhang:Evaluation:eUCRITEvsTink:init} ausfüllen. Danach wurden sie gebeten die Programmieraufgabe mit Tink und mit \zeierAPI zu lösen. Die Reihenfolge der beiden APIs wurde immer gewechselt. Nach den Programmieraufgaben sollten die Probanden jeweils den Fragebogen \ref{sec:Anhang:Evaluation:eUCRITEvsTink:nachAufgabe} ausfüllen. Wenn die Probanden die Programmeieraufgabe mit beiden APIs gelöst hatten, sollten sie zusätzlich den Fragebogen \ref{sec:Anhang:Evaluation:eUCRITEvsTink:letzter} ausfüllen. Die Fragebögen wurden mit einer LimeSurvey Installation online realisiert.

\subsection{Aufgabenstellung}\label{sec:Anhang:Evaluation:eUCRITEvsTink:aufgabe}
Die Aufgabenstellung und die Fragebögen der Studie.
\subsubsection{Introduction}

Thank you for taking part in our usability study about the usability of cryptographic APIs. 
During the study, you will implement the cryptographic functions inside a simple messenger application using two different Java APIs. The remaining code of the application will be provided to you. Afterwards, please fill in the online survey we prepared for you.

\subsubsection{Task Description}

The code provided to you shows a very simple messenger application. With this application, the users can exchange encrypted and signed messages. 
Your task is to implement the following methods: 
\begin{itemize}
    \item generateKey
    \item encrypt
    \item decrypt
    \item sign
    \item verify
\end{itemize}
You are allowed to change the method signatures to your needs. \\
\\
In addition, the generateKey() method should be implemented in a way that the key material is loaded after being generated once.\\
\\
The API that should be used is already included in the project. You will also be provided with a web browser showing the online documentation of the respective API. Please make use of this documentation as well (in addition to the Javadoc provided inside the IDE), since it is also part of our usability evaluation and should be rated by you in the end. \\
\\
We want your honest opinion about the usability of the APIs. Your voice will be recorded together with the screen capture. Thus, thinking aloud while programming will be the easiest way to tell us your opinion. Writing comments in your code is also fine. In addition, you will have the opportunity to share your thoughts within the survey. \\
\\
To test the application, just run the main-function and type a message into the lower window. Pressing send will then show you whether the task was solved correctly or not.

\subsection{Initialer Fragebogen}\label{sec:Anhang:Evaluation:eUCRITEvsTink:init}
Diese Fragen wurden zuerst gestellt, um den Probanden besser einschätzen zu können.
\begin{enumerate}
	\item Freitextfeld (Plichtfeld): Please ask the director of studies for your participant ID.
	\item Freitextfeld (Plichtfeld): How many years of programming experience do you have?
	\item Ja/Nein: Have you ever worked with a cryptographic API?
	\item Ja/Nein: Did you ever ask a question at a QuA Platform?
	\item Ja/Nein: Did you ever answer a question at a QuA Platform?
	\item Do you agree with these statements? \\
	Antwortmöglichkeiten: I strongly agree, I agree somewhat, I neither agree nor disagree, I disagree somewhat, I strongly disagree
	\begin{enumerate}
		\item I have a lot of experience in Java programming.
		\item I have a lot of cryptographic understanding.
	\end{enumerate}
	\end{enumerate}
Please solve the programming task with the \zeierAPI/Tink API.

\subsection{Fragebogen nach Programmieraufgabe}\label{sec:Anhang:Evaluation:eUCRITEvsTink:nachAufgabe}
Dieser Fragebogen wurde nach der \zeierAPI- und Tink-Programmieraufgabe ausgefüllt. Einige Fragen wurden aus dem von Acar \cite{acar_comparing_2017} entwickelten Fragenkatalog genommen um die Benutzbarkeit der APIs bestimmen zu können.
\begin{enumerate}[resume]
	\item Do you agree with these statements? \\
	Antwortmöglichkeiten: I strongly agree, I agree somewhat, I neither agree nor disagree, I disagree somewhat, I strongly disagree
	\begin{enumerate}
		\item \label{sec:Anhang:Evaluation:eUCRITEvsTink:nachAufgabe:FrageVier} The online documentation supported me in solving the task.
		\item \label{sec:Anhang:Evaluation:eUCRITEvsTink:nachAufgabe:FrageFuenf} I found my way easily through the online documentation.
		\item I had to understand how most of the assigned library works in order to complete the tasks.
		\item It would be easy and require only small changes to change parameters or configuration later without breaking my code.
		\item After doing these tasks, I think I have a good understanding of the assigned library overall.
		\item I only had to read a little of the documentation for the assigned library to understand the concepts that I needed for these tasks.
		\item The names of classes and methods in the assigned library corresponded well to the functions they provided.
		\item It was straightforward and easy to implement the given tasks using the assigned library.
		\item \label{sec:Anhang:Evaluation:eUCRITEvsTink:nachAufgabe:FrageEins} When I accessed the assigned library documentation, it was easy to find useful help.
		\item \label{sec:Anhang:Evaluation:eUCRITEvsTink:nachAufgabe:FrageZwei} In the documentation, I found helpful explanations.
		\item \label{sec:Anhang:Evaluation:eUCRITEvsTink:nachAufgabe:FrageDrei} In the documentation, I found helpful code examples.
		\item Using the information from the error message/ex\-ception, it was easy to fix my mistake.
		\item When I made a mistake, I got a meaningful error message/exception.
		\item I think the key generation was easy.
		\item I think the key storage was easy.
		\item I think encrypting and decrypting was easy.
		\item I think signing and verifying was easy.
	\end{enumerate}
	\item Schulnoten $1-6$: What grade would you give the eUCRITE API?
	\item \label{sec:Anhang:Evaluation:eUCRITEvsTink:nachAufgabe:FrageSechs} Schulnoten $1-6$: What grade would you give the eUCRITE API online Documentation?
	\item Freitext: What would you make better in the online documentation?
	\item Which algorithms and parameters did you use for solving the tasks? Why these ones?
	\end{enumerate}
Please solve the programming task with the \zeierAPI/Tink API.

\subsection{Letzter Fragebogen}\label{sec:Anhang:Evaluation:eUCRITEvsTink:letzter}
Dieser Fragebogen wurde ausgefüllt, wenn der Proband beide Programmieraufgaben gelöst hat.
\begin{enumerate}[resume]
	\item Freitext: Which advantages does the eUCRITE API have?
	\item Freitext: Which disadvantages does the eUCRITE API have?
	\item Freitext: Which advantages does the Tink API have?
	\item Freitext: Which disadvantages does the Tink API have?
\end{enumerate}
Thank you very much for your participation.

}{} 
%=================================================================================================================
\end{document}